\newif\iffigs\figstrue
\figsfalse

\documentstyle[12pt]{article}
\iffigs
  \input epsf
\else
\fi

\batchmode
\newfont{\footscrfont}{rsfs10}
  \newfont{\footbbbfont}{msbm10}
  \newfont{\manfont}{manfnt}
\errorstopmode

\newif\ifscrf\scrftrue
\ifx\footscrfont\nullfont
  \scrffalse
\fi

\newif\ifamsf\amsftrue
\ifx\footbbbfont\nullfont
  \amsffalse
\fi

\def\ppnumber{\vbox{\baselineskip14pt\hbox{CU-TP-896}
\hbox{hep-th/9805132 }}}
\def\ppdate{May 1997}
\def\pplogo{\vbox{\kern-\headheight\kern -15pt
\halign{##&##\hfil\cr&{
\ppnumber}\cr\rule{0pt}{2.5ex}&\ppdate\cr}
}}

\makeatletter
\date{}
\def\dedicatory#1{\def\@date{\normalsize\it#1}}
\def\subjclass#1{\def\@thefnmark{}\@footnotetext{1991
    {\it Mathematics Subject Classification.} #1}}
\def\keywords#1{\def\@thefnmark{}\@footnotetext{
    {\it Key words and phrases.} #1}}

\def\ps@firstpage{\ps@empty \def\@oddhead{\hss\pplogo}%
  \let\@evenhead\@oddhead 
}
\def\maketitle{\par
 \begingroup
 \def\thefootnote{\fnsymbol{footnote}}
 \def\@makefnmark{\hbox
 to 0pt{$^{\@thefnmark}$\hss}}
 \if@twocolumn
 \twocolumn[\@maketitle]
 \else \newpage
 \global\@topnum\z@ \@maketitle \fi\thispagestyle{firstpage}\@thanks
 \endgroup
 \setcounter{footnote}{0}
 \let\maketitle\relax
 \let\@maketitle\relax
 \gdef\@thanks{}\gdef\@author{}\gdef\@title{}\let\thanks\relax}

\def\abstract{\if@twocolumn
\section*{Abstract}
\else \small
\begin{center}
{\bf ABSTRACT}
\end{center}
\quotation
\fi}

\def\thebibliography#1{\section*{References\@mkboth
 {REFERENCES}{REFERENCES}}\small\list
 {[\arabic{enumi}]}{\settowidth\labelwidth{[#1]}\leftmargin\labelwidth
 \advance\leftmargin\labelsep
 \usecounter{enumi}}
 \def\newblock{\hskip .11em plus .33em minus .07em}
 \sloppy\clubpenalty4000\widowpenalty4000
 \sfcode`\.=1000\relax}

\newif\iffn\fnfalse

\@ifundefined{reset@font}{\let\reset@font\empty}{} 
\long\def\@footnotetext#1{\insert\footins{\reset@font\footnotesize
    \interlinepenalty\interfootnotelinepenalty
    \splittopskip\footnotesep
    \splitmaxdepth \dp\strutbox \floatingpenalty \@MM
    \hsize\columnwidth \@parboxrestore
   \edef\@currentlabel{\csname p@footnote\endcsname\@thefnmark}\@makefntext
    {\rule{\z@}{\footnotesep}\ignorespaces
      \fntrue#1\fnfalse\strut}}}

\makeatother

\ifamsf
  \newfont{\bigbbbfont}{msbm10 scaled\magstep2}
  \newfont{\bbbfont}{msbm10 scaled\magstep1}  
  \newfont{\smallbbbfont}{msbm8}
  \newfont{\tinybbbfont}{msbm6}
  \newfont{\smallfootbbbfont}{msbm7}
  \newfont{\tinyfootbbbfont}{msbm5}
  \newfont{\biggthfont}{eufm10 scaled\magstep2}
  \newfont{\gthfont}{eufm10 scaled\magstep1}  
  \newfont{\smallgthfont}{eufm8}
  \newfont{\tinygthfont}{eufm6}
  \newfont{\footgthfont}{eufm10}
  \newfont{\smallfootgthfont}{eufm7}
  \newfont{\tinyfootgthfont}{eufm5}
\fi

\ifscrf
  \newfont{\scrfont}{rsfs10 scaled\magstep1}  
  \newfont{\smallscrfont}{rsfs7}
  \newfont{\tinyscrfont}{rsfs7}
  \newfont{\smallfootscrfont}{rsfs7}
  \newfont{\tinyfootscrfont}{rsfs7}
\fi

\ifamsf
  \newcommand{\Bbb}[1]{\iffn
      \mathchoice{\mbox{\footbbbfont #1}}{\mbox{\footbbbfont #1}}
      {\mbox{\smallfootbbbfont #1}}{\mbox{\tinyfootbbbfont #1}}\else
      \mathchoice{\mbox{\bbbfont #1}}{\mbox{\bbbfont #1}}
      {\mbox{\smallbbbfont #1}}{\mbox{\tinybbbfont #1}}\fi}
  
\else
  \def\bigbbbfont{\bf}
  \def\Bbb{\bf}
  
\fi

\ifscrf
  \newcommand{\Scr}[1]{\iffn
    \mathchoice{\mbox{\footscrfont #1}}{\mbox{\footscrfont #1}}
    {\mbox{\smallfootscrfont #1}}{\mbox{\tinyfootscrfont #1}}\else
    \mathchoice{\mbox{\scrfont #1}}{\mbox{\scrfont #1}}
    {\mbox{\smallscrfont #1}}{\mbox{\tinyscrfont #1}}\fi}
\else
  \def\Scr{\cal}
\fi

\def\operatorname#1{\mathop{\rm #1}\nolimits}
\def\C{{\Bbb C}}

\def\P{{\Bbb P}}

\def\R{{\Bbb R}}
\def\Z{{\Bbb Z}}

\def\ad{\operatorname{ad}}

\def\bearray{\begin{eqnarray}}
\def\eearray{\end{eqnarray}}
\def\bearraynn{\begin{eqnarray*}}
\def\eearraynn{\end{eqnarray*}}
\def\bfig{\begin{figure}}
\def\efig{\end{figure}}

\def\opeq#1{\advance\lineskip#1 \advance\baselineskip#1
        \advance\lineskiplimit#1}

\def\cM{{\Scr M}}

\def\cD{{\Scr D}}

\def\cMc{{\hfuzz=100cm\hbox to 0pt{$\;\overline{\phantom{X}}$}\cM}}
\def\barcD{{\hfuzz=100cm\hbox to 0pt{$\;\overline{\phantom{X}}$}\cD}}

\ifamsf

\else

\fi

\def\ker{{\rm ker}}

\def\im{{\rm im}}
\def\dim{{\rm dim}}

\newtheorem{Proposition}{Proposition}[section]

\newtheorem{Theorem}{Theorem}[section]
\newtheorem{Lemma}{Lemma}[section]
\newtheorem{Corrolary}{Corrolary}[section]

\newcommand{\be}{\begin{equation}}
\newcommand{\ee}{\end{equation}}
\newcommand{\bea}{\begin{eqnarray}}
\newcommand{\eea}{\end{eqnarray}}
\newcommand{\bdm}{\begin{displaymath}}
\newcommand{\edm}{\end{displaymath}}
\newcommand{\nn}{\nonumber}
\newcommand{\bp}{\begin{Proposition}}
\newcommand{\ep}{\end{Proposition}}
\newcommand{\bt}{\begin{Theorem}}
\newcommand{\et}{\end{Theorem}}
\newcommand{\bl}{\begin{Lemma}}
\newcommand{\el}{\end{Lemma}}
\newcommand{\bc}{\begin{Corrolary}}
\newcommand{\ec}{\end{Corrolary}}

\begin{document}

\title{A noncommutative-geometric interpretation of the resolution of 
equivariant instanton moduli spaces}

\author{C.~I.~Lazaroiu$^{1,a}$}


\maketitle
\vbox{
\centerline{$^1$Department of Physics}
\centerline{Columbia University}
\centerline{N.Y., N.Y. 10027}
\medskip
\bigskip
}

\abstract{We generalize the recently proposed noncommutative ADHM construction
to the case of $\Gamma$-equivariant instantons over $\R^4$, with $\Gamma$ a 
Kleinian group. 
We show that a certain form of the inhomogeneous ADHM equations describes 
instantons over a noncommutative deformation of the Kleinian orbifold 
$\C^2/\Gamma$ and we discuss the relation of this with Nakajima's description 
of instantons over ALE spaces. In particular, we obtain a noncommutative 
interpretation of the minimal resolution of Kleinian singularities.}

\vskip .6in
$^a$ lazaroiu@phys.columbia.edu
\pagebreak

\section*{Introduction}

In a recent paper \cite{NS}, N. Nekrasov and A. Schwarz proposed 
a noncommutative version of the ADHM construction of instantons over 
$\R^4$. This can be obtained by deforming $\R^4$ to a noncommutative space 
and repeating  the steps of the usual ADHM construction in 
the new, noncommutative set-up. 

In the present paper we describe a generalization of these  
considerations to the equivariant case, i.e. to the case of instantons 
in V-bundles over an orbifold $\R^4/\Gamma$ with $\Gamma$ a discrete group. 
Since we are interested in making contact with the work of \cite{KN, N1, N2, 
N3,N4,N6} 
on the resolution of equivariant instanton moduli spaces,  
our discussion will 
be limited to Kleinian groups (i.e. to finite subgroups of $SU(2)$). 

Our description is essentially a $\Gamma$-equivariant 
version of the construction of \cite{NS}. While the commutative limit 
of \cite{NS} is the usual ADHM construction over $\R^4$, the commutative limit 
in our case will be an equivariant version of 
the ADHM construction, which was studied (in the case $\Gamma=\Z_n$) in 
\cite{FH, F, Au} and is implicit in the work of \cite{KN, N1, N2,N3,N4,N5,N6}.

Our results provide a re-interpretation of the quiver varieties of \cite{N2} 
as moduli spaces of equivariant noncommutative instantons. Moreover, they 
give a different version of the resolution mechanism, which is natural 
from the point of view of matrix theory. In fact, the motivation of the 
present work lies in an effort of understanding certain aspects of the  
resolution of Kleinian singularities by D-branes, a subject which was 
investigated in \cite{DM, JM}.

The plan of this paper is as follows. 
In section 1, we review equivariant instantons and the classical equivariant 
ADHM construction.
For lack of a suitable reference, we give a rather detailed account of the 
steps involved.  
In section 2, we develop its noncommutative version. 
To display the equivariance properties in a way that 
is easy to manipulate, we follow an approach which is more geometric than 
that adopted in \cite{NS}. Besides avoiding complicated matrix arguments, 
this has the advantage of illuminating the simple geometric origin of the 
construction, while clearly displaying the mathematical structures and 
technical assumptions involved. 
In section 3, we discuss the relation between our results and the work of 
\cite{KN,N1,N2}. Finally, section 4 presents our conclusions and suggestions 
for further research.

\section{The classical equivariant ADHM construction}

The well-known ADHM construction of instantons over $\R^4$ has a 
natural equivariant version \cite{FH, F, Au,N1}. 
The latter describes equivariant  
instantons over $\R^4$, or (equivalently) over $S^4$.

For our purpose it is convenient to 
formulate this  as follows. We start with a two dimensional complex hermitian 
vector space $Q$ and with a finite subgroup $\Gamma$ of $SU(Q)$. 
We let $Q_\R$ be the underlying real vector space of $Q$. 
The natural 
action of $\Gamma$ on $Q$ will be denoted by $\rho_Q$ (the fundamental 
representation of $\Gamma$). Choosing a real basis of $Q$ identifies it with 
$\R^4$, while the one -point compactification 
${\overline Q}=Q \cup \{\infty\} $ is identified with $S^4$. 
On the other hand, choosing a complex orthonormal basis of $Q$ identifies 
$Q$ with $\C^2$ and $SU(Q)$ with $SU(2)$. 
The  action of $\Gamma$ on $Q$ extends to the Alexandrov 
compactification ${\overline Q} \approx S^4$. The resulting 
action on ${\overline Q}$ has two fixed points, corresponding to 
$0 \in Q$ and to the point $\infty$.

Our instantons can be thought of as connections in a $V$-bundle 
(for background  on V-bundles, their topological invariants and 
equivariant connections, the reader is referred to \cite{Vbundles})
over $Q/\Gamma$ or (via Uhlenbeck's theorem on removable singularities) over 
${\overline Q}/\Gamma$. The extension to ${\overline Q}/\Gamma$ is essential 
for the topological classification of V-bundles, but the analytic aspects are 
clearer if one works on $Q/\Gamma$, and this paper we follow the 
latter approach. The only topological result we need is that 
$SU(r)$ $V$-bundles $E$ over $S^4$ are classified 
by their second Chern number $c_2(E)[S^4]$ and 
the isotropy representations $\rho_0,\rho_\infty$. 
The latter are defined as  the actions  
of $\Gamma$ on the fibers $E_0$, $E_\infty$ 
over the fixed points.

\subsection{V-bundles and equivariant connections over $Q/\Gamma$}

In this section we briefly recall some basic facts about hermitian V-bundles 
and their equivariant connections, as they apply to our situation.  

\

{\bf Generalities}

\

A hermitian V-bundle on $Q/\Gamma$ is a pair $(E,\phi)$, where $E$ is a 
hermitian vector bundle over $Q$ (of rank $r$) and $\phi$ is a family of 
unitary isomorphisms:
\bdm
\nn \phi_t(\gamma):E_t\rightarrow E_{\rho_Q(\gamma)t} 
\edm
($\gamma \in \Gamma$, $t \in Q$)
satisfying the compatibility conditions:
\bdm
\phi_t(\gamma_1\gamma_2)=\phi_{\rho_Q(\gamma_2)t}(\gamma_1)\phi_t(\gamma_2) \nn
\edm
A hermitian V-bundle isomorphism is a unitary bundle isomorphism compatible 
with this structure. 

The fundamental representation of $\Gamma$ induces an action $\nu$ by 
$*$-automorphisms of the algebra ${\cal F}(Q)$ of smooth functions defined 
over $Q$. $\nu(\gamma)f:=f^\gamma$ is given by:
\be
\label{nu}
f^\gamma(t)=f(\rho_Q(\gamma^{-1})t)
\ee
for all $t \in Q$. On ${\cal F}(Q)$ we consider the natural prehilbert module 
structure given by the hermitian form:
\bdm
<f,g>={\overline f}g~~ \nn,
\edm
which is $\nu$-equivariant in the sense:
\bdm
<f^\gamma, g^\gamma>=<f,g>^\gamma~~ \nn.
\edm 

Let ${\cal E}$ be the space of smooth sections of $E$. Then the 
$V$-structure of $E$ induces a representation $\nu_E$ by $\C$-linear 
automorphisms of ${\cal E}$ via $\nu_E(s):=s^\gamma$, with:
\bdm
s^\gamma(t):=\phi_{\rho_Q(\gamma^{-1})t}(\gamma)s(\rho_Q(\gamma^{-1})t)~~ \nn .
\edm
${\cal E}$ is an ${\cal F}(Q)$-bimodule, but $\nu_E$ is not an action by 
module automorphisms; rather, it is compatible with the action $\nu$ on 
${\cal F}(Q)$:
\bdm
(fs)^\gamma=f^\gamma s^\gamma~~ \nn .
\edm
The hermitian product of $E$ induces a hermitian metric $<,>$ on ${\cal E}$, 
which makes it into a hermitian ${\cal F}(Q)$-module 
(`prehilbert  module'). 
$\nu_E$ is also compatible with this structure
(we say that $\nu_E$ is `quasiunitary'):
\be
\label{quasiunitary}
<s_1^\gamma,s_2^\gamma>=<s_1,s_2>^\gamma~.
\ee

We also consider the linear action of $\Gamma$ on the space $X(Q)$ of 
vector fields over $Q$:
\bdm
X^\gamma:=\rho_Q(\gamma)_*~X \nn
\edm
for all $X \in X(M)$. This action is again compatible with $\nu$:
\bdm
(fX)^\gamma=f^\gamma X^\gamma~~ \nn .
\edm
If we pick a basis $(\epsilon_\alpha)_{\alpha=1..4}$ of $Q_\R$, then the 
vector 
fields $\partial_\alpha:=\frac{\partial}{\partial{\epsilon_\alpha}}$ form a 
basis of the ${\cal F}(Q)$-module $X(Q)$. It is easy to see that:
\bdm
\partial_\alpha^\gamma=\rho_Q(\gamma)_{\beta\alpha}\partial_\beta \nn
\edm
where $(\rho_Q(\gamma)_{\alpha\beta})_{\alpha, \beta=1..4}$ is the matrix of 
$\rho_Q(\gamma)$ in that basis. 

For any vector $t \in Q_\R$, the usual directional derivative 
$(\partial_tf)(s)=(d_sf)(t)$ gives a vector field $\partial_t \in X(Q)$. If 
$t=t^\alpha \epsilon_\alpha$, then $\partial_t=t^\alpha \partial_\alpha$. 
The vector space $T$ of such fields is isomorphic to $Q_\R$; it 
forms an abelian Lie algebra, which can be naturally identified with the Lie 
algebra of $(Q_\R, +)$. Then $(\partial_t)^\gamma=\partial_{t^\gamma}$, 
with $t^\gamma=\rho_Q(\gamma)t$.

\

\

\

{\bf Equivariant connections}

\

A connection $\nabla:X(Q)\times {\cal E}\rightarrow {\cal E}$ is 
{\em equivariant} if:
\be
\label{eq_conn}
\nabla_{X^\gamma}(s^\gamma)=(\nabla_X(s))^\gamma
\ee
for all $X \in X(M), s\in {\cal E}$ and $\gamma \in \Gamma$. 
Such connections are called {\em invariant} in 
\cite{FH,F,Au}. There one considered a more general set-up, in which 
equivariance cannot be realized in terms of a V-structure. 
In this sense, our  
situation (in which a V-bundle is given by hypothesis) 
is only a particular one, but it is this case 
which is related to  the work of \cite{KN}. 

We will always be interested in unitary connections, i.e.  connections 
compatible with the hermitian structure of $E$:
\be
\label{Ricci}
<\nabla_Xs_1,s_2>+<s_1,\nabla_Xs_2>=X<s_1,s_2>, ~\forall s_1, s_2 
\in {\cal E}, ~\forall X \in X(Q)~.
\ee
$\nabla$ is clearly determined by its restriction to $T$. 
Denoting $\nabla_{\partial_t}$ by $\nabla_t$, it therefore suffices to test 
(\ref{eq_conn},\ref{Ricci}) for the vector fields $\partial_t, t \in Q_\R$:
\be
\label{eq_conn_t}
\nabla_{t^\gamma}(s^\gamma)=(\nabla_t(s))^\gamma 
\ee
\be
\label{Ricci_t}
<\nabla_ts_1,s_2>+<s_1,\nabla_ts_2>=\partial_t<s_1,s_2>~~.
\ee

Since $Q$ is retractible, any vector bundle $E$ over $Q$ is 
topologically trivial; in particular, the module ${\cal E}$ is free. 
Fixing a unitary trivialization 
$E\approx Q\times \C^r$, we can define the trivial connection $\nabla^0$ 
on $E$ by $\nabla^0_X(s):=ds(X)$. Changing the trivialization affects 
$\nabla^0$ by a gauge transformation, so $\nabla^0$ is essentially 
unique. It is immediate that $\nabla^0$ is equivariant. 

A more explicit form of the equivariance condition (\ref{eq_conn}) can be 
obtained by choosing an orthonormal frame $(s_i)_{i=1..r}$ of $E$. Such a frame 
defines a unitary trivialization $\psi:{\underline \C}^r\rightarrow E$ via
$\psi_t(u_i):=s_i(t)$ ($t \in Q$), where 
${\underline \C}^r=Q\times \C^r$. Here $(u_i)_{i=1..r}$ is the canonical 
basis of $\C^r$. The sections $(s_i)_{i=1..r}$ give a basis of the free 
${\cal F}(Q)$-module ${\cal E}$. Since $\nu_E$ satisfies 
(\ref{quasiunitary}), $(s^\gamma_i)_{i=1..r}$ also 
form an orthonormal frame . Defining 
$\sigma_{\gamma,ij}$ by:
\bdm
s^\gamma_i=\sum_{j=1..r}{\sigma_{\gamma,ji}s_j}~~ \nn ,
\edm
we construct matrix-valued functions 
$\sigma_\gamma:=(\sigma_{\gamma,ij})_{i,j=1..r}$. They satisfy:
\be
\label{sigma_equiv}
\sigma_{\gamma_1 \gamma_2}=\sigma_{\gamma_1}(\sigma_{\gamma_2})^{\gamma_1}~.
\ee
Clearly $\sigma_\gamma(t)$ is unitary for all $t \in Q$. 
The connection 1-form $A=(A_{ij})_{i,j=1..r}$ in our frame is given by:
\be
\label{conn_matrix}
\nabla s_i=s_j A_{ji}~.
\ee 
It is not hard to see that the equivariance conditions (\ref{eq_conn}) are 
equivalent with:
\be
\label{eq_A}
\rho_Q(\gamma^{-1})^*A=\sigma_\gamma^{-1}(d+A)\sigma_\gamma
\ee
where $\rho_Q(\gamma^{-1})^*$ denotes the pull-back. This says that, under 
the action of $\Gamma$, $A$ transforms as a 1-form `up to gauge 
transformations'. Under a change of orthonormal frame 
(unitary gauge transformation) $s'_i=U_{ji} s_j$, $A$ and 
$\sigma_\gamma$ transform as:
\bea
\label{gauge_tf}
\nn A'=U^{-1}dU +U^{-1}AU  \\
\sigma'_\gamma=U^{-1}\sigma_\gamma U^\gamma, 
\eea
where $U:=(U_{ij})_{i,j=1..r}\in U(r)$.

$\psi$ can be used to transport the V-structure of 
$E$ to a V-structure $\phi_e$ on ${\underline \C^r}$ by requiring 
that the following diagrams commute:
\bdm
\begin{array}{ccc}
E_t & \stackrel{\phi_{E,t}(\gamma)}{\longrightarrow} & 
E_{\rho_Q(\gamma)t} \\
\psi_t ~\uparrow & \bigcirc & \uparrow ~\psi_{\rho_Q(\gamma)t}\\
\C^r & \stackrel{\phi_{e,t}(\gamma)}{\longrightarrow} & \C^r 
\end{array}~~ \nn .
\edm
This makes $\psi$ into an 
isomorphism of V-bundles. Identifying $\C^r$ with the space of column vectors, 
$\phi_{e,t}(\gamma)$ acts by matrix multiplication:
\bdm
\phi_{e,t}(\gamma)={\hat \phi}_{E,t}(\gamma)~\cdot \nn 
\edm
and ${\hat \phi}_{E,t}(\gamma)$ coincides with the 
matrix of $\phi_{E,t}(\gamma)$ taken in the bases $(s_i(t))_{i=1..r}$, 
$s_i(\rho_Q(\gamma)t)_{i=1..r}$ of $E_t$, respectively $E_{\rho_Q(\gamma)t}$. 
Moreover, we have:
\bdm
\sigma_{\gamma}(t)={\hat \phi}_{E,\rho_Q(\gamma^{-1})t}(\gamma) ~~ \nn .
\edm
This displays the way in which the V-structure of $E$ is encoded in the gauge 
transformations $\sigma_\gamma$. 

\

{\bf Product V-structures}

\

If $S$ is a hermitian $\Gamma$-module, construct an associated bundle 
${\underline S}:=Q\times S$ with the induced hermitian structure. The action 
$\rho_S$ of $\Gamma$ on $S$ induces a unitary $V$-structure on 
${\underline S}$ via:
\bdm
\phi_{S,t}(\gamma):=\rho_S(\gamma) , ~\forall t \in Q, 
~\forall \gamma \in \Gamma 
~~ \nn .
\edm
Such hermitian V-bundles will be called {\em product} V-bundles. They are 
the natural analogues of trivial bundles. In this case, the action $\nu_S$ 
is simply: 
\be
\label{nu_S}
\nu_S:=\rho_Q\otimes_{\C} \nu~~.
\ee
In fact, the free module ${\cal S}$ has a distinguished presentation:
\bdm
{\cal S}=S\otimes_{\C} {\cal F}(Q)~~ \nn .
\edm
Note that, if $L$ is the trivial $\Gamma$-module (of dimension 1), then 
${\cal L}={\cal F}(Q)$ and $\nu_L=\nu$. 

It is easy to see that a hermitian V-bundle $E$ is isomorphic with a 
product hermitian V-bundle 
iff $E$ admits an orthonormal frame $(s_i)_{i=1..r}$ in which all of the 
matrices $\sigma_\gamma$ are constant. In this case, ${\sigma_\gamma}$ form 
a unitary representation $\sigma$ of $\Gamma$, and $E$ is isomorphic to 
the product 
V-bundle $Q\times \Sigma$, where $\Sigma=(\C^r,\sigma)$ is the corresponding 
hermitian $\Gamma$-module.

\subsection{The equivariant ADHM construction}

The construction proceeds as follows. One starts 
with two hermitian $\Gamma$ -modules 
$V$, $W$, of dimensions $v, w$ and with {\em equivariant} ADHM data 
${\cal B} \in 
Hom_\Gamma(V, Q\otimes V)$, $I \in Hom_\Gamma(W,\Lambda^2 Q\otimes V)$, 
$j \in Hom_\Gamma(V,W)$ 

To display the equivariance properties, 
the more traditional data $B_1, B_2 \in Hom(V,V)$ have been replaced  
by ${\cal B} \in Hom_\Gamma(V, Q\otimes V)$. 
$B_1, B_2$ can be recovered by choosing any orthonormal basis 
$(e_1,e_2)$ of $Q$ and decomposing 
${\cal B}=e_1\otimes B_1 + e_2 \otimes B_2$. Then the equivariance of  
${\cal B}$ is expressed by:
\bdm 
\rho_Q(\gamma)e_1\otimes \rho_V(\gamma)B_1\rho_V(\gamma^{-1})+
\rho_Q(\gamma)e_2\otimes \rho_V(\gamma)B_2\rho_V(\gamma^{-1})=
e_1\otimes B_1 +e_2\otimes B_2 \nn 
\edm
(for all $\gamma \in \Gamma$). 
Writing 
$\rho_Q(\gamma)=\left( \begin{array}{cc}
\gamma_1 & \gamma_2 \\
-{\overline \gamma}_2 & {\overline \gamma_1} 
\end{array}\right) \in SU(2)$ for the matrix of $\gamma$ in 
our basis, 
this reads:
\bea
\rho_V(\gamma^{-1})B_1\rho_V(\gamma)=\gamma_1B_1 +\gamma_2 B_2 \nn \\
\rho_V(\gamma^{-1})B_2 \rho_V(\gamma)=-{\overline \gamma}_2 B_1 + 
{\overline \gamma}_1 B_2~~ \nn .
\eea
The basis $e_1,e_2$ gives a nonzero vector $e_1\wedge e_2$ of the 
1-dimensional complex vector space $\Lambda^2 Q$; this allows us to identify 
$\Lambda^2 Q$ with $\C$. Under this identification, $I$ becomes an element 
$i$ of $Hom(W,V)$. Note that the representation of $\Gamma$ induced on 
$\Lambda^2 Q$ is trivial. Therefore, as far as equivariance properties are 
concerned, we can always view $I$ as the element $i$ of 
$Hom(W,V)$. Indeed, we will often do this below. 
With this identification, the equivariance conditions on $i,j$ are:
\bea
\rho_W(\gamma^{-1}) i \rho_V(\gamma)=i \nn \\
\rho_V(\gamma^{-1})j \rho_W(\gamma)=j~~ \nn .
\eea

Let $P_\Gamma(V,W)=Hom_\Gamma(V, Q\otimes V)\oplus Hom_\Gamma(W,V)\oplus 
Hom_\Gamma(V,W)$ be the total space of ADHM data. This space admits a natural 
action of $G_V:=U_\Gamma(V)$ given by :
\bdm 
(B_1, B_2, i, j) \rightarrow (UB_1U^{-1}, UB_2U^{-1}, Ui, jU^{-1})~~ \nn .
\edm
Here $U_\Gamma(V)$ is the set of all unitary transformations of $V$ which 
commute with the representation $\rho_V$. 
$P_\Gamma$ can be endowed 
with a natural structure of hermitian quaternionic vector space, given  
by the complex structures $I,J,K$, where $I$ is 
the original  complex structure  of $P_\Gamma$, $J$ is given by :
\bdm
J((e_1\otimes B_1+e_2\otimes B_2)\oplus i\oplus j):=
(-e_1\otimes B_2^+ + e_2\otimes B_1) \oplus (-j^+)\oplus (i^+) \nn 
\edm
and $K=IJ$. 

With this quaternionic structure, the action of 
$G_V$ is unitary symplectic. 
The associated hyperkahler moment map is given by the usual expressions:

\bea
\mu_r(B_1,B_2,i,j)=\frac{i}{2}([B_1,B_1^+]+[B_2,B_2^+]+ii^+ -j^+j) \nn \\
\label{moment2} 
\mu_c(B_1,B_2,i,j)=[B_1,B_2]+ij~~ \nn .
\eea

To describe instantons (as opposed to more general objects), one 
requires that the stabilizer of $({\cal B},i, j)$ in $G_V$ be trivial. 
The set of all data of trivial stabilizer in $G_V$  
will be denoted by $P^{reg}_\Gamma(V,W)$.

In the classical construction, the ADHM data are also 
required to satisfy the homogeneous equations:
\be
\label{ADHM1}
\mu_r(B_1,B_2,i,j)=0
\ee
(the real ADHM equation)
\be
\label{ADHM2}
\mu_c(B_1,B_2,i,j)=0
\ee
(the complex ADHM equation).

The main claim of the ADHM  
construction is that the hyperkahler quotient 
${\cal M}^{reg}_0(V,W):=
\{({\cal B},i,j) \in P^{reg}_\Gamma(V,W )|\mu_r({\cal B},i,j)=
\mu_c({\cal B},i,j)=0\}/G_V $ is isomorphic with the moduli space
of equivariant SU(w)- instantons (framed at $\infty$) 
over $S^4$ of second Chern number $c_2(V)[S^4]=\dim V$ 
and isotropy representations $\rho_0=\rho_{Q\otimes V \oplus W \ominus 
(V \oplus V)}, \rho_\infty=\rho_W$. 

To obtain the associated instanton connection, one defines maps :
\bea
\label{st}
\sigma_t:=({\cal B}-  t\otimes) \oplus j :V \rightarrow Q\otimes V \oplus W 
\nn \\
\tau_t:=[{\cal B}\wedge_Q-id_V \otimes (t\wedge )]+i:Q\otimes V \oplus W 
\rightarrow \Lambda^2Q \otimes V
\eea
where the operator ${\cal B}\wedge_Q:Q\otimes V\rightarrow \Lambda^2Q 
\otimes V$ is given by:
\bdm
{\cal B}\wedge_Q(s \otimes v):={\cal B}(v)\wedge_Q s \nn 
\edm
(for all $s \in Q$ and $v \in V$). 
Here $\wedge_Q$ is the wedge product in $Q$. 
It is easy to see that:
\bea
(t\otimes ) \circ \rho_V(\gamma)=\rho_{Q\otimes V}(\gamma) \circ 
[(\rho_Q(\gamma^{-1})t)\otimes ] \nn \\
(t\wedge )\circ \rho_{Q\otimes V}(\gamma)=\rho_V(\gamma)\circ 
[(\rho_Q(\gamma^{-1})t) \wedge ]~~ \nn .
\eea
Together with the equivariance properties of the ADHM data, this gives:
\bea
\label{eq_st}
\sigma_t \circ \rho_V(\gamma)=\rho_{Q\otimes V \oplus W}(\gamma)\circ 
\sigma_{\rho_Q(\gamma^{-1})t} \nn \\
\tau_t\circ \rho_{Q\otimes V \oplus W}(\gamma)=\rho_V(\gamma)\circ 
\tau_{\rho_Q(\gamma^{-1})t}~.
\eea
Using the basis $(e_1,e_2)$ of $Q$ 
to identify $Q\otimes V \approx V\oplus V$ by 
$u=v_1\otimes e_1+v_2 \otimes e_2 \in Q \otimes V \rightarrow (v_1,v_2) \in 
V \oplus V$ and $\Lambda^2 Q \approx \C$ gives the familiar expressions:
\bea
\sigma_t:=\left(\begin{array}{c}
B_1 - t_1 \\
B_2 -t_2 \\
j
\end{array}\right) \in Hom(V, V\oplus V \oplus W) \nn \\
\tau_t:=\left(\begin{array}{ccc}
-B_2+t_2 & B_1-t_1 & i
\end{array}\right) \in Hom(V\oplus V\oplus W, V) \nn 
\eea
for all $t=t_1~e_1+ t_2~e_2 \in Q$. 

For data satisfying $\mu_c({\cal B},i,j)=0$ {\em and} $\mu_r({\cal B},i,j)=0$, 
the trivial stabilizer requirement is equivalent with the pair of conditions:

(a)If $T$ is a $\Gamma$-invariant subspace of $V$ such that 
$B_1(T) \subset T$ , $B_2(T) \subset T$ and $T \subset \ker j$, then we have 
$T=0$

(b) If $T$ is a $\Gamma$-invariant 
subspace of $V$ such that $B_1(T) \subset T$ , 
$B_2(T) \subset T$ and $\im i \subset T$, then we have $T=V$, 

\noindent and with the pair of (apparently stronger) conditions obtained 
from these by dropping the requirement of $\Gamma$-invariance of $T$. 

On the other hand, if $\mu_c({\cal B},i,j)=0$, then  
(a) is equivalent to 
injectivity of $\sigma_t$ for all $t \in Q$, while (b) is equivalent 
to surjectivity of $\tau_t$  for all $t \in Q$. 
Moreover, the complex ADHM equation implies that 
$ \tau_t \sigma_t=0$ for all $t$.
These last statements are true  
regardless of the value of $\mu_r({\cal B},i,j)$. 
 
Thus for data satisfying both (\ref{ADHM1}) and (\ref{ADHM2}), 
we have monads 
\bdm
0 \longrightarrow V \stackrel{\sigma_t}{\longrightarrow} Q \otimes V \oplus W 
\stackrel{\tau_t}{\longrightarrow} V \longrightarrow 0 \nn 
\edm
for all $t \in Q$. These give a vector bundle monad  over $Q$:
\be
\label{bundle_monad}
0 \longrightarrow {\underline V} \stackrel{\sigma}{\longrightarrow} 
{\underline Q} \otimes {\underline V} \oplus {\underline W} 
\stackrel{\tau}{\longrightarrow} {\underline V} \longrightarrow 0 ~~,
\ee
and the  equivariance properties imply that $\sigma, \tau$ are V-bundle 
morphisms. 

The instanton bundle is given by the cohomology of (\ref{bundle_monad}):
\bdm
E:=\ker \tau / \im \sigma~~ \nn .
\edm
To induce an instanton connection on $E$, one proceeds as follows. 
First, define the operator: 
\be
\label{D}
{\cal D}_t :=(\tau_t^+, \sigma_t) \in Hom( V\oplus V, Q\otimes V \oplus W)
\ee
Then $E_t$ can be identified with $\ker {\cal D}^+_t$,  
which represents $E$ as a subbundle of the bundle 
$U:={\underline Q}\otimes {\underline V}\oplus {\underline W}$. 
 
Next, the trivial connection $\nabla ^0$ on $U$  
induces a connection on $E$ by:
\bdm
\nabla_X(s):={\cal P}\nabla^0_X(s), ~\forall X \in X(M), 
~\forall s \in {\cal E} \nn 
\edm
where we let $P$ be the 
orthogonal projector in the hermitian bundle $U$ on the 
fibers of the subbundle $E$ and ${\cal P}$ the associated projector on 
sections. This connection turns out \cite{DK} 
to be anti-self-dual and of square-integrable curvature; it is the desired 
instanton.

The equivariance properties (\ref{eq_st}) give:
\be
\label{equiv_D}
\rho_{V\oplus V}(\gamma)\circ {\cal D}^+_t=
{\cal D}^+_{\rho_Q(\gamma)t}\circ \rho_{Q\otimes V\oplus W}(\gamma)~,
\ee
which says that ${\cal D}^+$ is a V-bundle morphism. 
It follows that the map 
$\phi_{Q\otimes V \oplus W,t}=\rho_{Q\otimes V\oplus W}(\gamma)$
induces a unitary isomorphism $\phi_{E,t}(\gamma)$ 
between 
the fibers $E_t$ and $E_{\rho_Q(\gamma)t}$. These identifications define 
the V-bundle structure of $E$. For $t=0$ we obtain the isotropy  
representation $\rho_0=\rho_{Q\otimes V\oplus W}(\gamma)|_{E_0}$. 
Since ${\cal D}_0^+$ is an isomorphism on the orthogonal complement of $E_0$, 
we have $E_0=Q\otimes V \oplus W \ominus (V \oplus V)$ as an abstract 
$\Gamma$-module. 
On the other hand, taking 
$t \rightarrow \infty$, 
shows that the isotropy representation 
at $\infty$ is given by $E_\infty=W$. 

${\cal P}$ is an orthogonal projector of  
the hermitian ${\cal F}(Q)$-module ${\cal U}$,  
whose image coincides with the space of sections 
${\cal E}$ of $E$. Property (\ref{equiv_D}) can be rewritten :
\be
\label{equiv_P}
{\cal P}\circ \nu_U=\nu_U \circ {\cal P}~.
\ee
Since the trivial connection on $U$ is  
equivariant, (\ref{equiv_P}) implies the equivariance of $\nabla$.

An explicit expression of $\nabla$  can be obtained by picking  
a unitary trivialization (`a gauge') of $E$, i.e. a map: 
\bdm 
\psi: Q \rightarrow Hom(\C^w, Q\otimes V \oplus W) \nn 
\edm
such that $\psi_t^+\psi_t=1_{\C^w}$ and ${\cal D}^+_t\psi_t=0$. By noting that 
$\psi_t(u_i)$ gives an orthonormal basis of $E_t$ for all $t \in Q$, 
one sees that the connection 1-form $A$ of $\nabla$ coincides with the matrix 
of the (form-valued) linear operator: 
\bdm
{\overline A}:=\psi^+d\psi \nn 
\edm
in the canonical basis 
$(u_i)_{i=1..w}$ of $\C^w$. 
Equivariance of $\nabla$ can be formulated in terms of the connection 
1-form $A$ as in (\ref{eq_A}), and a trivial computation shows that 
$\sigma_\gamma(t)$ coincides with the matrix of the linear operator:
\bdm
{\overline \sigma}_\gamma(t) :=\psi_t^+ \rho_{Q\otimes V \oplus W}(\gamma) 
\psi_t \nn 
\edm
in the canonical basis of $\C^w$.

\subsection{Quiver description of the equivariant ADHM data}

To make contact with the quiver description of \cite{KN,N2}, let 
$R_i$ ($i=0..r$) denote the irreps of $\Gamma$ (with $R_0$ the trivial irrep.) 
and $R$ be its regular representation. Let $n_i:=\dim_\C R_i$ 
and $n:=\dim_\C R$. We have $R=\oplus_{i=0..r}\C^{n_i}\otimes R_i$. 
Let $V=\oplus_{i=0..r}{V_i \otimes R_i}$, $W=\oplus_{i=0..r}{W_i \otimes R_i}$ 
be the decompositions of $V,W$ in irreps of $\Gamma$. Define a 
symmetric matrix $A=(a_{ij})_{i,j=0..r}$ by:
\bdm
Q\otimes R_i =\oplus_{j=0..r}{\C^{a_{ij}}\otimes R_j} ~~ \nn ,
\edm
and let $C=2I-A$. 
The McKay correspondence asserts that $C$ is the extended Cartan matrix 
of a simply-laced Lie algebra ${\bf g}_\Gamma$. Then $A$ is the 
incidence matrix of the associated extended Dynkin diagram $\Delta$ 
(in particular, $a_{ij} =a_{ji} \in \{0,1\}$). 

Schur's lemma gives decompositions:
\bea
Hom_\Gamma(V, Q\otimes V)=\oplus_{i,j=0,~a_{ij}=1}{Hom(V_i,V_j)} \nn \\
Hom_\Gamma(W,V)=\oplus_{i=0..r}{Hom(W_i,V_i)} \nn \\
Hom_\Gamma(V,W)=\oplus_{i=0..r}{Hom(W_i,V_i)}~~ \nn .
\eea
Accordingly, we have decompositions:
\bea
{\cal B}=\oplus_{i,j=0..r,~a_{ij}=1}{B_{ij}} \nn \\
i=\oplus_{k=0..r}{i_k} \nn \\
j=\oplus_{k=0..r}{j_k}~~ \nn .
\eea
Viewing $B_{ij},B_{ji}$(for $a_{ij}=1$) as associated to the links of $\Delta$ 
and $i_k,j_k$ as associated to its nodes recovers the desired quiver 
description. Define:
\bea
{\vec v}:=(v_0...v_r)^t \nn \\
{\vec w}:=(w_0...w_r)^t \nn
\eea
where $v_i:=\dim_\C V_i$ and $w_i:=\dim_\C W_i$. Then the isotropy 
representations at $0,\infty$ are given by:
\bea
E_0=Q\otimes V \oplus W -(V\oplus V)=\sum_{i=0..r}{u_i R_i} \nn \\
E_\infty=W=\oplus _{i=1..r}{w_iR_i} \nn
\eea
with $u_i=w_i-\sum_{j=0..r}{c_{ij}v_j}$. Since $ker C$ is one-dimensional 
(being spanned by the vector ${\vec n}=(n_0..n_r)$ associated to the regular 
representation), knowledge of $\rho_0, \rho_\infty$ determines ${\vec w}$ 
completely, but fixes ${\vec v}$ only up to a multiple of ${\vec n}$. 

\subsection{Ideal instantons}

A partial compactification of ${\cal M}_0^{reg}(V,W)$ is given by the moduli 
space of ideal equivariant instantons:
\bdm
{\cal M}_0(V,W):=\{({\cal B},i,j) \in P_\Gamma(V,W)| \mu_r({\cal B},i,j)=
\mu_c({\cal B},i,j)=0\}/G_V~~,
\edm
which is obtained by dropping the trivial stabilizer condition. 
Since the action of $G_V$ is no longer free, the ideal instanton moduli space 
is, in general, singular. 

As explained for example in \cite{DK}, ideal instantons are pairs formed 
by a usual instanton (of lower second Chern number) and a set of points 
(which may be void).  
The latter are sometimes called `small instantons' in the physics literature 
and are obtained as follows. 
If $({\cal B},i,j)$ satisfies the homogeneous ADHM equations but has 
nontrivial stabilizer in $G_V$, then one of the conditions (a), (b) above 
fails, so that there will exist a (finite) set of 
points $t_1..t_s \in Q$ such that ${\cal D}_{t_i}^+$ is 
not surjective (i.e. either $\sigma_{t_i}$ is not injective, either 
$\tau_{t_i}$ is not surjective). 
At such points, the dimension of $E_t$ jumps, so that the 
bundle $E$ is replaced by a sheaf. If one attempts to apply the ADHM 
construction nonetheless, one will obtain a connection having singularities 
at these points. Intuitively, a part of the instanton has collapsed to zero 
size. In general, the dimension of $E_t$ may jump by more than one, and in 
this case the point $t$ is taken with multiplicity equal to this jump. Such 
points correspond to `coalescing small instantons'. Since they are directly 
related to ADHM data with a nontrivial stabilizer, small instantons are 
responsible for the singularities of ${\cal M}_0(V,W)$. 

In our case, the equivariance properties of $\sigma, \tau$ show that:
\bea
\label{small_i}
\rho_V(\gamma)ker \sigma_t=ker \sigma_{\rho_Q(\gamma)t} \nn \\
\rho_V(\gamma)im \tau_t = im\tau_{\rho_Q(\gamma)t}
\eea
so that the sets $S_\sigma$, $S_\tau$ of points where 
$\sigma, \tau$ fail to have maximal rank are $\Gamma$-invariant 
(including their multiplicities). Hence the set $S_\sigma  \cup S_\tau$ of 
`small instantons' (with multiplicity) also has this property. Clearly the 
situation is different according to whether $t=0$ (the fixed point of the 
action of $\Gamma$ on $Q$) or $t$ belongs to the set $Q-\{0\}$, on which 
$\Gamma$ acts freely. In the latter case, we have 
$|\Gamma |$ different `images' and one can  deduce a constraint on $V$. 
Indeed, it not hard to see that small instantons can occur outside of the 
orbifold point only if $V$ contains the regular representation of $\Gamma$ as 
a direct summand. 

Therefore, if $V$ does not contain $R$, then small instantons can only occur 
at the fixed point (in this case, $ker \sigma_0$ and $ker \tau_0$ must be 
invariant subspaces of $V$, which constrains the allowed multiplicities). 
Hence for such choices of $V$, the singularities of ${\cal M}_0(V,W)$ are a 
direct consequence of the presence of the fixed point.

\section{Noncommutative version}

\subsection{Intuitive considerations} 

Fix $\zeta <0$. In this section, we wish to repeat the ADHM construction by 
starting from equivariant data $({\cal B},i,j) \in P_\Gamma(V,W)$ 
which obey the following inhomogeneous form of the ADHM equations:
\bea
\label{ADHM_inhom}
\mu_r(B_1,B_2,i,j)=-\frac{i}{2}\zeta\otimes Id_{V} \nn \\
\mu_c(B_1,B_2,i,j)=0~~.
\eea
We do not impose any extra-conditions on $({\cal B},i,j)$. 

In the classical construction, anti-self-duality of the ADHM connection is 
a consequence of the relations $\tau\sigma=0, \sigma^+ \sigma=\tau\tau^+$, 
which assure that ${\cal D}_t^+{\cal D}_t=1_2 \otimes \Delta$, 
with $\Delta$ a positive operator. These relations  
are induced by the {\em homogeneous} equations (\ref{ADHM1}),
(\ref{ADHM2}). 
The beautiful observation of \cite{NS} 
is that these key properties can also be satisfied for data obeying 
(\ref{ADHM_inhom}) if one promotes
$t_1, t_2$ to non-commuting variables $z_1, z_2$ such that:
\bea
\label{ct}
[z_1,z_1^+] + [z_2, z_2^+]=\zeta ; & [z_1,z_2] = 0 ~~. 
\eea
Then the steps of the usual ADHM construction can be repeated in this 
noncommutative set-up. 

The conditions (\ref{ct}) are solved by 
taking $[z_i,z_j^+]=2 \eta\delta_{ij}$, with $\eta=\frac{\zeta}{4}<0$.  
Since the fundamental action of $\Gamma$ preserves these relations, it 
follows that $\rho_Q$ extends to a 
morphism of the $*$-algebra generated by $z_i$, allowing us to define 
a noncommutative deformation of the orbifold $Q/\Gamma$ and to carry over the 
classical equivariant arguments to the noncommutative case. 

We now proceed to give a more rigorous form to 
these intuitive ideas. For this, we must give a clear description of the 
noncommutative deformation of the base space and develop the noncommutative 
analogues of the building blocks which entered in the equivariant form of the 
classical construction.

\subsection{The noncommutative version of the base space}
  
The desired noncommutative deformation of $Q$ can be achieved in two 
different but equivalent ways. The first method 
(which we will follow here) is inspired by the ideas of deformation 
quantization while the second approach proceeds via Weyl systems and 
is explained in Appendix 1.

\

{\bf  Approach via deformation quantization}

\

By virtue of the Gelfand-Naimark theorem, the commutative space $Q_\R$ can 
be described by its (non-unital) $\C^*$-algebra
$C_0(Q_\R):=A_0$ of continuous functions vanishing at infinity. 
For our purpose  it is more convenient to start with the algebra 
$C_u(Q_\R)$ of bounded uniformly continuous functions on $Q_\R$, 
which has a unit. 

To define a noncommutative deformation controlled by the 
parameter $\eta$, view $Q_\R$ 
as a symplectic space with  the symplectic form $\omega:=Im<,>$. 
Consider the natural (strongly-continuous) action $\alpha$ of $(Q_\R,+)$ by 
$*$-automorphisms of  
$C_u(Q_\R)$ given by translations $\alpha_u:f\rightarrow f_u$, where:
\be
\label{trsl}
f_u(t):=f(t+u) 
\ee
for all $t \in Q_\R$. 
The set of smooth (i.e. ${\cal C}^\infty$) vectors for this action 
is the subalgebra ${\cal B}^\infty_0$  of infinitely differentiable functions 
over $Q_\R$, whose partial derivatives of any order 
are bounded on $Q_\R$. The deformation quantization of ${\cal B}^\infty_0$ 
along the symplectic form 
$\eta \omega$ can be achieved \cite{Rieffel} by replacing the original 
product of ${\cal B}^\infty_0$ by the Moyal product, which is formally given 
by:
\be
\label{Moyal}
(f\times_\eta g)(t):= [e^{-\frac{i}{2}\eta\pi^{\mu\nu}\partial_u\partial_v}
f(u)g(v)]_{u=v:=t}
\ee
(here $\pi^{\mu\nu}$ is the inverse of the matrix $\omega_{\mu\nu}$). 
This gives a new involutive algebra structure on the set ${\cal B}^\infty_0$, 
which we denote by ${\cal B}^\infty_\eta$. 
Note that the {\em sets} ${\cal B}^\infty_0$ and 
${\cal B}^\infty_\eta$ coincide, and the involution of 
${\cal B}^\infty_\eta$ is still given by the usual complex conjugation. 
As explained in \cite{Rieffel}, 
one can  introduce a norm $|| . ||_\eta$ on ${\cal B}^\infty_\eta$, compatible 
with the deformed product $\times_\eta$. Taking the completion of 
${\cal B}^\infty_\eta$ with respect to this norm gives a (unital) 
$C^*$ -algebra $B_u^\eta$. 
It turns out \cite{Rieffel} that $\alpha$ extends to an action by 
$*$-automorphisms of the new algebra structure 
$(B_u^\eta,\times_\eta,+,||.||_\eta)$, and the 
subalgebra of smooth vectors for this action coincides with 
${\cal B}^\infty_\eta$.

For our purpose, the completion $B_u^\eta$ will be of little import, since 
we will concentrate mainly on the geometric aspects of the problem. 
For this, it will suffice to consider the restricted formalism of 
gauge connections developed in \cite{Connes1, Connes2}, which 
involves only the subalgebra ${\cal B}^\infty_\eta$. 
To summarize, then, ${\cal B}^\infty_\eta$ is the set ${\cal B}^\infty_0$,  
with the usual addition and complex conjugation of 
functions, but with the new product $\times_\eta$ given by (\ref{Moyal}). 
The action $\alpha$ of $(Q_\R,+)$ on ${\cal B}^\infty_\eta$ is the usual 
action by translations (\ref{trsl}), which is an algebra morphism of 
${\cal B}^\infty_\eta$:
\bdm
\alpha_u(f\times_\eta g)=\alpha_u(f)\times_\eta \alpha_u(g) \nn ~~.
\edm 

Since  $\alpha$ acts by $*$-automorphisms of  
${\cal B}^\infty_\eta$, one obtains an action 
$\ad_\alpha$ of $(Q_\R,+)$ by derivations, compatible with the involution.  
Clearly $\ad_\alpha(t)(f)$ is the usual directional derivative $\partial_tf$ 
and the derivation property reads:
\bdm
\partial_t(f\times_\eta g)=(\partial_tf)\times_\eta g+ f \times_\eta 
(\partial_tg) ~~ \nn .
\edm
For later use, define the differentiation operator 
$d:{\cal B}^\infty_\eta \rightarrow (Q_\R)^*\otimes {\cal B}^\infty_\eta$ by:
\bdm
(df)(t):=\partial_tf ~~ \nn .
\edm
Here $(Q_\R)^*:=Hom_\R(Q_\R, \R)$. 

The action (\ref{nu}) of $\Gamma$ on ${\cal F}(Q_\R)$  
restricts to an action by  *-automorphisms of ${\cal B}^\infty_0$. 
It turns out that  $\nu(\gamma)$ are also  
$*$-automorphisms of the deformed algebra structure ${\cal B}^\infty_\eta$ 
for any $\eta >0$, so that we have:
\bdm
(f\times_\eta g)^\gamma=f^\gamma\times_\eta g^\gamma \nn 
\edm
(this is shown in Appendix 1). 
The pair $({\cal B}^\infty_\eta,\nu)$ is 
the natural noncommutative generalization of the orbifold $Q/\Gamma$ and 
will henceforth be called a `noncommutative orbifold'. 

For part  of the following,  we will further restrict to the 
*-subalgebra 
${\cal A}_\eta$ of ${\cal B}^\infty_\eta$ whose underlying set is the 
Schwarz space ${\cal S}(Q_\R)$. As discussed in \cite{Rieffel}, 
${\cal A}_\eta$ is not only a subalgebra, but also a bilateral ideal of 
${\cal B}_\eta^\infty$. 
The reason for considering ${\cal A}_\eta$ is that we wish  
the operators:
\begin{eqnarray}
\label{zs}
z_i\times_\eta =(z_i+\eta\partial_{{\overline z}_i}) \nn\\
{\overline z}_i\times_\eta =({\overline z}_i-\eta\partial_{z_i})
\end{eqnarray}
to preserve our modules. Indeed, these operators preserve ${\cal A}_\eta$, 
but not ${\cal B}^\infty_\eta$. 
Note that ${\cal A}_\eta$ does not have a unit.  
As we explain  below, if one restricting to 
${\cal A}_\eta$-modules only leads to some unpleasant 
features of the formalism. Since 
${\cal B}_\eta^\infty$ and ${\cal A}_\eta^\infty$ contain only smooth vectors 
for the action 
(\ref{trsl}), the formalism of \cite{Connes1} is applicable to both 
of them. However, nonunitality of ${\cal A}_\eta$ obstructs   
a standard formulation of connection matrices and matrix unitary gauge 
transformations, due to lack of a good analogue of orthonormal bases of 
the associated hermitian modules. 
We believe that a better way of treating this 
problem is to work with the {\em unital} algebra 
${\cal B}^\infty_\eta$ from 
the outset, in which case a more natural formalism of connections can be 
recovered; this is the approach we will propose in this paper. In Appendix 1 
we sketch how this can be technically achieved, provided that a certain 
`regularity' condition holds. 
Note, however, 
that one can always construct a noncommutative connection {\em operator} 
even if one uses 
${\cal A}_\eta$-modules only (provided that the assumptions needed for the  
argument of \cite{NS} hold). The regularity condition of Appendix 2 
is needed only if one wishes to recover a standard matrix formalism.

\subsection{Noncommutative hermitian V-bundles}

\

{\bf Generalities} 

\

The considerations of section 1  suggest the following definition: 
A `noncommutative hermitian V-bundle' over $({\cal B}^\infty_\eta,\nu)$ 
is a pair $({\cal E}, \nu_E)$ where ${\cal E}$ is 
a finite projective {\em right} hermitian ${\cal B}^\infty_\eta$-module and 
$\nu_E$ an action of 
$\Gamma$ on ${\cal E}$ by $\C$-linear automorphisms, satisfying the 
compatibility conditions:
\bea
\nu_E(sf)=\nu_E(s)\nu(f) \nn \\
<\nu_E(\gamma)(s_1), \nu_E(\gamma)(s_2)>=\nu(\gamma)(<s_1,s_2>)~~\nn .
\eea
We will denote $\nu_E(\gamma)(s)$ by $s^\gamma$. 
An isomorphism of noncommutative hermitian V-bundles is a unitary module 
isomorphism compatible with this action.

        We will be interested in unitary connections as defined in 
\cite{Connes1}, i.e. in maps:
\bdm
\nabla:{\cal E}\rightarrow (Q_\R)^* \otimes {\cal E} \nn 
\edm
satisfying the `Leibniz' property:
\bdm
\nabla_t(sf)=\nabla_t(s)f + s\partial_t(f) ~,\forall s \in {\cal E}, ~\forall 
f \in {\cal B}_\eta^\infty, ~\forall t \in Q_\R \nn 
\edm
where $\nabla_t s:=(\nabla s)(t)$ as usual, and the `Ricci property'  
(\ref{Ricci_t}). 
Such a connection will be called {\em equivariant} if it also satisfies 
condition (\ref{eq_conn_t}). Clearly the map $\nabla_t$ obtained by 
restricting a usual equivariant connection $\nabla_X$ to the space of vector 
fields $T$ defined in section 1 is a noncommutative equivariant connection in 
the above sense. Therefore, we have a generalization of the usual notion of 
equivariant connections. 

Given a right ${\cal B}_\eta^\infty$-module ${\cal E}$ and a connection 
$\nabla$,  an associated 
${\cal A}_\eta$-module with a connection is obtained  by restriction of 
scalars. We use the same notations ${\cal E}$, $\nabla$ for the latter.

\

{\bf Free noncommutative V-bundles}

\

If the ${\cal B}_\eta^\infty$-module ${\cal E}$ is free (of rank r), 
then we can give a more 
concrete description of our objects, as in the commutative case. 
For this, choose an orthonormal basis  
$(s_i)_{i=1..r}$ of ${\cal E}$. Quasiunitarity (\ref{quasiunitary}) of 
$\nu_E$ allows us to define unitary ${\cal B}^\infty_\eta$ - valued matrices 
$\sigma_\gamma:=(\sigma_{{\gamma}, ij})_{i,j=1..r} \in 
Mat(r,{\cal B}^\infty_\eta)$ by:
\bdm
\nu_E(\gamma)(s_i)=\sum_{j=1..r}{s_j\sigma_{\gamma, ji}} \nn 
\edm
These satisfy (\ref{sigma_equiv}) and encode the noncommutative V-structure of 
${\cal E}$. The connection 1-form is again defined by (\ref{conn_matrix}); 
it is an element of $(Q_\R)^*\otimes Mat(r,{\cal B}_\eta^\infty)$ or, 
equivalently, 
an $\R$-linear map $A:Q_\R \rightarrow  Mat(r,{\cal B}^\infty_\eta)$. Its 
values are `anti-hermitian' matrices. 
A simple 
computation shows that the equivariance condition is again equivalent to 
(\ref{eq_A}), if we define $\rho_Q(\gamma^{-1})^*A$ by:
\bdm
(\rho_Q(\gamma^{-1})^*A)(t):=A(\rho_Q(\gamma^{-1})t)^\gamma, 
~\forall t \in Q_\R ~~ \nn . 
\edm
This reduces to the usual pull-back in the commutative case. 
Under a change of orthonormal basis $s'_i:=s_j U_{ji}$ of ${\cal E}$, we again 
have the transformations (\ref{gauge_tf}), where now 
$U:=(U_{ji})_{i,j=1..r} \in U(r,{\cal B}^\infty_\eta)$. The curvature 
$F$ of our connection is an element of $\Lambda^2(Q_\R)^*\otimes 
Hom_{{\cal B}^\infty_\eta}({\cal E}, {\cal E})$, i.e. a skew-symmetric map 
$F:Q_\R \times Q_\R \rightarrow 
Hom_{{\cal B}^\infty_\eta}({\cal E}, {\cal E})$, 
defined by:
\bdm
F(s,t)\sigma:=(\nabla_s \nabla_t -\nabla_t \nabla_s)\sigma, 
~\forall \sigma \in {\cal E} ~~ \nn .
\edm 
The matrix of $F(s,t)$ in the basis $s_i$ is:
\bdm
F(s,t)=\partial_sA(t)-\partial_tA(s)+ [A(s), A(t)] ~~ \nn .
\edm
Choosing an orthonormal basis $(\epsilon_{\alpha})_{\alpha=1...4}$ of 
$Q_\R$ and defining $F_{\alpha\beta}:=F(\epsilon_{\alpha},\epsilon_{\beta})$ 
recovers the standard formula:
\bdm
F_{\alpha\beta}=\partial_\alpha A_\beta -\partial_\beta A_\alpha + 
[A_\alpha, A_\beta]_\times \nn 
\edm
(where the commutator $[A_\alpha, A_\beta]_\times$ is computed with the 
Moyal product).
Finally, the frame $(s_i)_{i=1..r}$ gives a right-linear module isomorphism 
$\psi:\C^r \otimes {\cal B}^\infty_\eta \rightarrow {\cal E}$, defined by 
$\psi(u_i):=s_i$, where $(u_i)_{i=1..r}$ is the canonical basis of $\C^r$, 
viewed as a basis of $\C^r \otimes {\cal B}^\infty_\eta$ via 
$u_i \equiv u_i \otimes 1_{{\cal B}^\infty_\eta}$. 
Then it is easy to repeat the 
considerations of section 1 , regarding $\psi$, in our noncommutative context.

The existence of a unit of ${\cal B}_\eta^\infty$ is crucial for 
being able to define orthonormal bases $(s_i)_{i=1..r}$ of 
${\cal B}_\eta^\infty$-modules. Indeed, the orthonormality condition 
$<s_i,s_j>=\delta_{ij}1_{{\cal B}_\eta^\infty}$ pressuposes a unit in the 
base algebra. Therefore, if  one  considers only  
${\cal A}_\eta$ -modules ${\cal E}$, one is  
not able to define such bases. In this case, it is only possible 
to define the 
connection form in a more general basis of ${\cal E}$; as a result, its 
values will not be hermitian matrices.  
Moreover, unitary automorphisms of ${\cal E}$ will no 
longer be represented by `unitary' matrices. 
Therefore, in this situation one is not, in general, able to recover a nice 
noncommutative analogue of the usual matrix formalism.

\

{\bf Noncommutative product V-bundles}

\

The bundles appearing in (\ref{D}) are product V-bundles.  
This allows for a straightforward definition of their noncommutative 
deformation. In the commutative case, the space of sections ${\cal S}$ of 
${\underline S}$ has a distinguished presentation 
as the space of smooth maps from $Q$ into the vector space $S$, that is, we 
have a module isomorphism ${\cal S}=S\otimes_{\C} {\cal B}^\infty_0$
(from now on, in the commutative case, we restrict scalars from 
${\cal F}(Q_\R)$ to ${\cal B}_0^\infty$). 
The noncommutative deformation of this is simply:
${\cal S}^\infty_\eta:=S\otimes_{\C} {\cal B}^\infty_\eta$, 
which has a natural {\em bimodule} structure over ${\cal B}^\infty_\eta$.  
As a $\C$-vector space, ${\cal S}^\infty_\eta$ coincides with ${\cal S}$, 
but the multiplications 
$sf, fs$ for $f \in {\cal B}^\infty_\eta$ and 
$s \in {\cal S}$ are now computed on components by 
replacing the usual point-wise product with the Moyal product $\times_\eta$. 
The hermitian product on ${\cal S}^\infty_\eta$ is induced from ${\cal S}$, 
and has the desired bilinearity properties with respect to the new 
module structure. If $S_1,S_2$ are two hermitian vector spaces, 
and ${\cal S}^{\infty,\eta}_i:=S_i\otimes_{\C} {\cal B}^\infty_\eta $ the 
associated modules, then we can define the tensor product 
$\otimes_\eta:=\otimes_{{\cal B}_\eta^\infty}$ of modules  
${\cal S}_1^{\infty,\eta}\otimes_\eta {\cal S}_2^{\infty,\eta}$ over 
${\cal B}_\eta^\infty$ by 
considering the left module structure on ${\cal S}_1^{\infty,\eta}$ and the right 
module structure on ${\cal S}_2^{\infty,\eta}$. Since 
${\cal S}_i^{\infty,\eta}$ are 
${\cal B}_\eta^\infty$-bimodules, the result is an ${\cal B}_\eta^\infty$-bimodule. 
If $s_i \in {\cal S}_i^{\infty,\eta}$ are sections of ${\underline S}_i$, then 
$s_1\otimes_\eta s_2$ is computed by taking the usual tensor product 
on the vector space parts $S_i$ and the Moyal product $\times_\eta$ on 
the components. Since the Moyal product is associative, the tensor 
product $\otimes_\eta$ is associative as well. We have 
$(S_1\otimes S_2)\otimes_\C {\cal B}_\eta^\infty=({\cal S}_1^{\infty,\eta})\otimes_\eta 
{\cal S}_2^{\infty,\eta}$. 

(\ref{nu_S}) is a quasiunitary $\C$-linear action on  
${\cal S}^\infty_\eta$, for {\em any} 
value of $\eta$, and gives a noncommutative V-bundle structure to  
${\cal S}^\infty_\eta$. 
If $S_1, S_2$ are hermitian $\Gamma$-modules, then the actions on 
the modules $(S_1\oplus S_2)\otimes_{\C} {\cal B}_\eta^\infty$ and 
$(S_1\oplus S_2)\otimes_{\C} {\cal B}_\eta^\infty=({\cal S}_1^{\infty,\eta})\otimes_\eta 
{\cal S}_2^{\infty,\eta}$ are given by 
$\nu_{S_1\oplus S_2}=\nu_{S_1}\oplus \nu_{S_2}$ and 
$\nu_{S_1\otimes S_2}:=\nu_{S_1}\otimes_\eta \nu_{S_2}$ respectively.

It is easy to see that a noncommutative V-bundle ${\cal E}$ is 
isomorphic 
to a product V-bundle iff ${\cal E}$ is free and it admits an orthonormal basis
in which the matrices $\sigma_\gamma$ are constant:
\bdm
\alpha_u(\sigma_\gamma)=\sigma_\gamma, ~\forall u \in Q \nn 
\edm
In this case, $\sigma_\gamma$  give a unitary representation  of 
$\Gamma$, and ${\cal E}$ is isomorphic with the product 
V-bundle $\Sigma \otimes_{\C} {\cal B}_\eta^\infty$ 
induced by the $\Gamma$-module $\Sigma:=(\C^r, \sigma)$. 

Finally, given a ${\cal B}^\infty_\eta$-module 
$S_\eta=S\otimes {\cal B}^\infty_\eta$, we can consider the 
${\cal B}_\eta^\infty$-submodule 
$S_\eta:=S\otimes {\cal A}_\eta=S_\eta^\infty\otimes_{{\cal B}_\eta^\infty}
{\cal A}_\eta$, where the bilateral ideal ${\cal A}_\eta$ 
of ${\cal B}_\eta^\infty$ is viewed as a 
${\cal B}_\eta^\infty$-bimodule. In the following, we will denote such 
submodules simply by $S\otimes {\cal A}_\eta$, but it is understood that they 
are always to be viewed as ${\cal B}_\eta^\infty$-modules. 
For any two sections $s_1,s_2$ of $S_\eta$, 
we have $<s_1,s_2> \in {\cal A}_\eta$. 
In particular, no ${\cal B}^\infty$-submodule of such a module admits 
orthonormal bases. Note that the modules $S\otimes {\cal A}_\eta$ are not free 
over ${\cal B}_\eta^\infty$.

\subsection{${\cal D}, {\cal D}^+$ as operators on sections}

To reinterpret the ADHM construction in the language of noncommutative 
geometry, we view $\sigma, \tau, {\cal D}, {\cal D}^+$  as operators on 
sections. To achieve this, we define a `tautological' section  $z$ of 
${\underline Q}$, whose value at any point $t\in Q$ is given by $t$. 
Then the section form of our operators is obtained by replacing 
$t\otimes, t\wedge$ in the definition of $\sigma, \tau$ with the point-wise 
tensor/wedge products of sections 
$z\otimes_0, z\wedge_0$:
\bea
\label{st_0}
\sigma:=({\cal B}- z\otimes_0) \oplus j \nn \\
\tau:=[{\cal B}\wedge_Q-id_V \otimes (z\wedge_0 )]+i ~.
\eea
We obtain ${\cal B}^\infty_0$-right linear operators:
\bea
\sigma:V\otimes {\cal A}_0 \rightarrow (Q\otimes V \oplus W)\otimes {\cal A}_0 \nn \\
\tau:(Q\otimes V \oplus W)\otimes {\cal A}_0 \rightarrow V\otimes {\cal A}_0 
\nn \\
{\cal D}: (V\oplus V) \otimes {\cal A}_0\rightarrow (Q\otimes V \oplus W)
\otimes {\cal A}_0 \nn \\
{\cal D}^+:(Q\otimes V \oplus W)\otimes {\cal A}_0  
\rightarrow (V\oplus V) \otimes {\cal A}_0 \nn
\eea
Note that $z\otimes_0, z\wedge_0$ preserve ${\cal B}_0^\infty$-modules of 
the form $S\otimes {\cal A}_0$, but not modules of the form 
$S \otimes {\cal B}^\infty_0$. This is the 
reason for restricting to submodules of the form $S\otimes {\cal A}_0$. 

If $\nu_Q$ is the action of $\Gamma$ on the space 
${\cal Q}=Q\otimes {\cal A}_0$ of sections of ${\underline Q}$, then 
$z$ obeys the trivial equivariance law $\nu_Q(\gamma)(z)=z$. 
This implies:
\bea
\label{z_eq}
(z\otimes_0)\circ \nu_V(\gamma)=\nu_{Q\otimes V}(\gamma)\circ (z \otimes_0 )
\nn \\
(z \wedge_0 ) \circ \nu_{Q\otimes V}(\gamma)=\nu_V(\gamma)\circ (z\wedge_0 )
\eea
which gives the following reformulation of the equivariance properties: 
\bea
\label{D_eq}
\sigma\circ \nu_V(\gamma)=\nu_{Q\otimes V \oplus W}(\gamma) \circ \sigma \nn \\
\tau \circ \nu_{Q\otimes V \oplus W}(\gamma)=\nu_V(\gamma) \circ \tau \\
{\cal D}\circ\nu_{V\oplus V}(\gamma)=\nu_{Q\otimes V \oplus W}(\gamma)\circ 
{\cal D} \nn \\
{\cal D}^+\circ\nu_{Q\otimes V\oplus W}(\gamma)=\nu_{V \oplus V}(\gamma)\circ 
{\cal D}^+ \nn
~.
\eea

\subsubsection{The noncommutative equivariant ADHM construction}

Let $\eta :=\zeta/2$. 
Then we define ${\cal B}^\infty_\eta$-right linear operators: 
\bea
\sigma:V\otimes {\cal A}_\eta \rightarrow (Q\otimes V \oplus W)
\otimes {\cal A}_\eta \nn \\
\tau:(Q\otimes V \oplus W)\otimes {\cal A}_\eta \rightarrow V\otimes 
{\cal A}_\eta \nn \\
{\cal D}: (V\oplus V) \otimes {\cal A}_\eta\rightarrow (Q\otimes V \oplus W)
\otimes{\cal A}_\eta \nn \\
{\cal D}^+:(Q\otimes V \oplus W)\otimes {\cal A}_\eta 
\rightarrow (V\oplus V) \otimes {\cal A}_\eta \nn
\eea
by replacing $z\otimes, z\wedge$ in (\ref{st_0}) with $z\otimes_\eta, 
z\wedge_\eta$:
\bea
\sigma:=({\cal B}- z\otimes_\eta) \oplus j \nn \\
\tau:=[{\cal B}\wedge_Q-id_V \otimes (z\wedge_\eta )]+i ~~ \nn .
\eea
Here $z\otimes_\eta$, $z\wedge_\eta$ are defined by using the usual tensor 
product on the vector space parts and the Moyal product on the scalar 
components.These operators 
are {\em right} ${\cal B}_\eta^\infty$-linear. 
Note that, since $z$ does not belong to 
$Q\otimes {\cal B}_\eta^\infty$, this is not quite the module tensor product 
consider above (despite the similar notation). However, it clearly has the 
same equivariance properties.

As remarked in \cite{NS}, the commutation relations obeyed by 
$z_i, {\overline z}_i$ assure that the validity of the key properties  
$\tau\sigma=0, \sigma^+ \sigma=\tau\tau^+$. Thus we have:
\be
\label{diag}
{\cal D}^+{\cal D}=1_2 \otimes \Delta= 
\left(\begin{array}{cc} 
\Delta & 0 \\
0 & \Delta \end{array}\right)
\ee
with $\Delta:V \otimes {\cal A}_\eta \rightarrow V \otimes {\cal A}_\eta$. 

We define a 
{\em right} ${\cal B}^\infty_\eta$-module by 
${\cal E}_\eta:=\ker {\cal D}^+$; this is a submodule of 
${\cal U}_\eta:=(Q\otimes V \oplus W)\otimes{\cal A}_\eta$. 
In the commutative limit $\eta=0$, 
${\cal E}_\eta$ reduces to the module of sections ${\cal E}$ 
of the bundle $E$ constructed before.
Note that ${\cal E}_\eta$ does {\em not}, in general, 
coincide with ${\cal E}$ as a set.
${\cal E}_\eta$ carries an ${\cal A}_\eta$-valued hermitian product 
induced from ${\cal U}_\eta$. 

The equivariance 
properties of $z,{\cal D}, {\cal D}^+$ are again given by (\ref{z_eq}), 
(with $\otimes_0$ replaced by $\otimes_\eta$) and (\ref{D_eq}). 
It follows that ${\cal E}_\eta$  is $\Gamma$-invariant. 
In particular, ${\cal E}_\eta$ carries a natural action $\nu_{{\cal E}_\eta}$ 
of $\Gamma$ by $\C$-linear quasiunitary automorphisms, 
given by the restriction of $\nu_U$.  
Therefore, $({\cal E}_\eta,\nu_E)$ is a hermitian noncommutative 
V-bundle over $({\cal A}_\eta, \nu)$. 

By {\em assuming}
that $\Delta$ is invertible on 
$V \otimes {\cal A}_\eta$, it follows that the operator:
\be
\label{Proj}
{\cal P}_\eta:=1-{\cal D}(1_2\otimes \Delta^{-1}){\cal D}^+
\ee
gives an orthogonal projector of the hermitian module ${\cal U}_\eta$ onto 
${\cal E}_\eta$. 
Then the null connection $\nabla^0=d$ of 
${\cal U}_\eta$ induces a connection in ${\cal E}_\eta$:
\bdm
\nabla_t(s):={\cal P}_\eta \nabla^0_t(s), ~\forall t \in Q_\R, ~\forall s 
\in {\cal E}_\eta~~ \nn .
\edm 
It is  easy to see that the usual argument \cite{DK} for anti-self-duality of 
the induced connection carries over unaffected to the noncommutative case. 
Therefore, the connection $\nabla_t$ has anti-self-dual curvature. 
This is our noncommutative instanton. Indeed, the curvature $F$ of $\nabla$ is 
given by the `Gauss equation':
\bdm
<F(v,w)s_1,s_2>=<\Pi_ws_1, \Pi_vs_2>-<\Pi_vs_1, \Pi_ws_2>~~ \nn ,
\edm
where $\Pi=(1-{\cal P}_\eta)d$ and $\Pi_v=(1-{\cal P}_\eta)\partial_v$. Since 
${\cal D}^+s=0$ for all $s \in {\cal E}_\eta$, we have 
${\cal D}^+d=(d{\cal D}^+)=(d\xi)^+\pi_{Q\otimes V}$ (on ${\cal E}_\eta$), 
where $ \pi_{Q\otimes V}:(Q\otimes V \oplus W)\otimes {\cal A}_\eta
\rightarrow (Q\otimes V)\otimes {\cal A}_\eta$ is 
the orthogonal projection and :
\bdm
\xi:=\left(\begin{array}{cc}
-{\overline z}_2 & z_1 \\
{\overline z}_1 & z_2
\end{array}\right) \nn
\edm
with $z_i:Q\rightarrow \C$ the coordinate functions in an orthonormal basis 
of $Q$. This gives $\Pi|_{{\cal E}_\eta}={\cal D}(1_2 \otimes \Delta^{-1})
(d\xi)^+\pi_{Q\otimes V}|_{{\cal E}_\eta}$, which implies:
\bdm
F(v,w)={\cal P}_\eta [\pi_{Q\otimes V}^+ 
d\xi(v)(1_2\otimes \Delta)d\xi(w)^+\pi_{Q\otimes V} -(v <--> w)~] \nn 
\edm
i.e. 
\be
\label{curvature}
F={\cal P}_\eta\pi_{Q\otimes W}^+\left(\begin{array}{cc}
d{\overline z}_2\frac{1}{\Delta}dz_2+dz_1\frac{1}{\Delta}d{\overline z}_1 &
-d{\overline z}_2\frac{1}{\Delta}dz_1+dz_1\frac{1}{\Delta}d{\overline z}_2 \\
-d{\overline z}_1\frac{1}{\Delta}dz_2+dz_2\frac{1}{\Delta}d{\overline z}_1 & 
d{\overline z}_1\frac{1}{\Delta}dz_1+dz_2\frac{1}{\Delta}d{\overline z}_2
\end{array}\right)\pi_{Q\otimes W}~~.
\ee

$\Gamma$-invariance of ${\cal E}_\eta$ reads:
\bdm
\nu_U{\cal P}_\eta={\cal P}_\eta \nu_U, 
\edm
which (together with equivariance of $\nabla^0$) 
immediately yields equivariance of $\nabla$. 

At this point, we have constructed a noncommutative instanton connection 
in the ${\cal B}_\eta^\infty$-module ${\cal E}_\eta$. Since this module 
cannot admit orthonormal bases, we do not yet have a nice analogue of 
the usual matrix formalism. Assuming as in \cite{NS} that ${\cal E}_\eta$ 
is free as an  ${\cal A}_\eta$-module, 
one can consider orthogonal bases 
of the ${\cal A}_\eta$-module obtained from ${\cal E}_\eta$ by restriction of 
scalars and define connection matrices in such bases.  
If one attempts to construct the  unitary isomorphism 
$\psi:\C^w \otimes_{{\cal A}_\eta} {\cal A}_\eta\rightarrow {\cal E}_\eta$ 
of \cite{NS} directly at this level, one runs into various problems of 
normalization which make the formalism rather cumbersome. 
Instead of attempting to develop a formalism along these lines, 
we take a different point of view, arguing for the existence of an extension 
of the projector ${\cal P}_\eta$ to the module 
${\cal U}_\eta^\infty$, which will 
allow us to construct an extension ${\cal E}_\eta^\infty$ of the 
${\cal B}_\eta^\infty$-module ${\cal E}_\eta$. Provided that certain technical 
assumptions are satisfied, it is then possible to argue that the 
{\em extended} ${\cal B}_\eta^\infty$-module ${\cal E}_\eta^\infty$ is free, 
and in this case one can consider orthonormal bases of it, thereby obtaining 
a satisfactory formalism of connection forms. In the following, we will only 
sketch the idea of this construction. A more detailed account is given in 
Appendix 1, in the framework of Weyl quantization.

\

{\bf Extension to ${\cal E}_\eta^\infty$}

\

Although the operators ${\cal D}$, ${\cal D}^+$ do not preserve 
the ${\cal B}_\eta^\infty$-modules $(V \oplus V )\otimes {\cal B}_\eta^\infty$ 
and $(Q\otimes V \oplus W)\otimes {\cal B}_\eta^\infty$, they are always 
defined on these modules if we allow them to take values in a more general 
space (essentially a space of distributions). 
Thus we can always define 
${\cal E}_\eta^\infty$ to be the kernel of 
${\cal D}^+$ on $(Q\otimes V \oplus W)\otimes {\cal B}_\eta^\infty$, 
which is a right ${\cal B}_\eta^\infty$-module. 
In particular, relation (\ref{diag}) still holds for these generalized 
operators. If we make the (nontrivial) 
assumption that the operator ${\cal D}(1_2\otimes \Delta^{-1}){\cal D}^+$ 
extends to an operator which preserves the space 
${\cal U}_\eta^\infty:=(Q\otimes V \oplus W)\otimes {\cal B}_\eta^\infty$, 
then (\ref{Proj}) gives an orthogonal projector ${\cal P}_\eta^\infty$ of 
${\cal U}_\eta^\infty$, which extends the projector 
${\cal P}_\eta$ and maps ${\cal U}_\eta^\infty$ onto 
${\cal E}_\eta^\infty$. Using ${\cal P}_\eta^\infty$ instead 
of ${\cal P}_\eta$ and starting with the null connection on 
${\cal U}_\eta^\infty$, we have 
an induced connection on ${\cal E}_\eta^\infty$ which extends 
the connection $\nabla$. The self-duality argument proceeds unaffected and 
the curvature is still given by (\ref{curvature}). 

By further  assuming 
that there exists a unitary isomorphism 
$\psi:\C^w \otimes {\cal B}^\infty_\eta \rightarrow {\cal E}^\infty_\eta$ 
(which amounts to saying that 
${\cal E}^\infty_\eta$ is free as a ${\cal B}^\infty_\eta$-module), we can 
construct ${\cal B}_\eta^\infty$-valued matrices $A, \sigma_\gamma$ as before. 
Again a trivial computation shows that:
\bea
A_{ji}=<u_j,(\psi^+ \circ d \circ \psi) (u_i)> \nn \\
\sigma_{\gamma,ji} = <u_j,(\psi^+ \circ \nu_U(\gamma) \circ\psi) (u_i)> ~~ 
\nn .
\eea
where $u_i\equiv u_i\otimes 1_{{\cal B}_\eta^\infty}$. 
The construction of $\psi$ is further discussed in Appendix 1.

To give at least a rough justification for the extendibility of 
${\cal P}_\eta$, 
note that, in the commutative case (and for ADHM data of trivial stabilizer), 
the projector ${\cal P}$ of section 2 
preserves the module ${\cal U}_0^\infty$. Indeed, 
$P_t \in Hom(U,U)$ is a smooth function of $t \in Q$ and has a limit at 
infinity given by the orthogonal projector
$P_\infty$ on the fiber $E_\infty=W$. 
Since ${\cal P}^\infty_\eta$ should  
`continuously' reduce (in a suitable sense) to ${\cal P}$ as $\eta$ 
tends to zero, this indicates that the desired condition may be 
satisfied. Further discussion of how the extension of  
${\cal P}_\eta$ could be achieved is given in Appendix 1.

\section{Discussion of the moduli space}

Our noncommutative instanton moduli space:
\be
\label{qv}
{\cal M}_{(\zeta,0)}(V,W):=
\{({\cal B},i,j)\in P_\Gamma(V,W)|\mu_r({\cal B},i,j)=
-\frac{i}{2}\zeta Id_V; \mu_c({\cal B},i,j)=0 \}/G_V \nn 
\ee
is a particular example of a {\em quiver variety}. Such varieties were 
studied in detail in \cite{N2}, to which we refer the reader for 
background. For the convenience of the non-expert reader, 
the main results of relevance for us are summarized in Appendix 2. 

As expected, the limit $\zeta=0$ gives the moduli space ${\cal M}_0(V,W)$ of 
$\Gamma$-equivariant {\em ideal} instantons on $Q$, 
which was discussed in section 1. In the case $\zeta <0$, the results of 
\cite{N2} imply that ${\cal M}_{(\zeta,0)}(V,W)$ is always smooth and 
that it provides a resolution of singularities of ${\cal M}_0(V,W)$.  
The differential and complex structures of ${\cal M}_{(\zeta,0)}(V,W)$ 
are  independent of the parameter $\zeta$, which controls only 
its Kahler structure.  
Our construction gives a noncommutative-geometric interpretation of 
this resolution: as in \cite{NS}, 
deforming the orbifold to its noncommutative version 
resolves the singularities of the ideal instanton moduli space. However, 
as we discuss below, the geometry of the orbifold situation is richer. 
The reason is that the noncommutative deformation effectively eliminates the 
orbifold point, thus implicitly achieving the resolution of the orbifold. 

As reviewed in Appendix 2, a general quiver variety is obtained by 
considering arbitrary central levels 
${\vec \zeta}$ of both the real and the complex moment maps in (\ref{qv}). 
For generic values of ${\vec \zeta}$, such a variety is a smooth hyperkahler 
manifold. 
For fixed $V$ and $W$, the differential structure of a generic 
(in the sense defined in Appendix 2) quiver variety 
is independent of the value of ${\vec \zeta}$, which controls only its 
hyperkahler structure. Since diagonal levels as in (\ref{qv}) are always 
generic, it follows that ${\cal M}_{(\zeta,0)}(V,W)$ gives a differentiable 
model for all generic quiver varieties of the same type $(V,W)$. Therefore, 
working with our particular deformation of the ADHM equations suffices 
to capture all of the {\em differential}-geometric aspects of the resolution.  
The situation is more complicated if one takes into account the complex 
structure, or the full hyperkahler structure of ${\cal M}_{(\zeta,0)}(V,W)$, 
as we discuss below.

\subsection{Instanton interpretation}

In certain situations, the {\em hyperkahler} manifold 
${\cal M}_{(\zeta,0)}(V,W)$ admits an interpretation as a moduli space of 
instantons over a smooth ALE space. Such moduli spaces were constructed in 
\cite{KN}. They are also given by quiver varieties, but not any quiver 
variety admits such an interpretation. The obstruction to this is a 
`tracelessness' condition, discussed in detail in Appendix 2, which can be 
satisfied by our moment map level only if the $\Gamma$-module $V$ is such that
there exists an $i \in \{0..r\}$ for which $V_i=0$.

If this condition is satisfied, one can in general find 
a smooth ALE space $X_{(\xi, 0)}$ such that ${\cal M}_{(\zeta,0)}(V,W)$
coincides (as a {\em hyperkahler} manifold) with the moduli space of 
instantons over $X_{(\xi, 0)}$ (the parameter $\xi$ is related to $\zeta$
as explained in Appendix 2). The underlying complex 
manifold of $X_{(\xi, 0)}$ is the minimal resolution of the Kleinian orbifold 
$Q/\Gamma$, while $\xi$ controls its Kahler structure. 
Therefore, in this case, the resolution 
mechanism admits a purely classical geometric description, namely as a blow-up 
of the orbifold $Q/\Gamma$ to $X_{(\xi,0)}$, which takes equivariant ideal  
instantons over $Q/\Gamma$ into instantons over the resolved space. 

The reason why the blow-up of $Q/\Gamma$ suffices (in this case) to resolve 
the equivariant ideal instanton moduli space can be understood as follows. 
When $V_i=0$ for some $i$, the $\Gamma$ -module $V$ does not 
contain the regular representation. Therefore, as we discussed in section 1,
point-like equivariant instantons can only occur at the orbifold point. 
The resolution of $Q/\Gamma$ removes this point,  
and, since $V$ does not contain $R$, Proposition 9.2. of \cite{KN} assures 
us that {\em no} point-like instantons 
can form on the resolved space $X_{(\xi,0)}$. 

A particularly pleasant outcome of this situation is that we can obtain the 
ALE space $X_{(\xi,0)}$ itself via our noncommutative construction. Indeed, 
it was shown in \cite{KN,N2} that 
$X_{(\xi,0)}={\cal M}_{(\zeta, 0)}(R\ominus R_0, Q)$, i.e. $X_{(\xi,0)}$ 
coincides with a particular moduli space of SU(2) instantons over itself. 
By our construction, $X_{(\xi,0)}$  
also coincides with the 
moduli space of noncommutative equivariant instantons over the 
deformation of the {\em orbifold} $Q/\Gamma$. The limit $\xi\rightarrow 0$ 
($\zeta \rightarrow 0$) 
gives the moduli space of equivariant ideal instantons 
over $Q/\Gamma$, of $c_2(E)[S^4]=|\Gamma|-1$ and 
isotropy representations $E_0=2R_0$ (trivial 2-dimensional representation) 
and $E_\infty=Q$, which indeed coincides \cite{N2} with $X_0=Q/\Gamma$.

\subsection{Sheaf-theoretic interpretation}

The underlying complex manifold of ${\cal M}_{(\zeta,0)}(V,W)$ also admits 
an interpretation as a moduli space of equivariant torsion-free 
sheaves (framed over the line at infinity) over $\C\P^2$. 
We give a sketch of the relevant arguments in Appendix 2.

\subsection{Methodological considerations}

The instanton interpretation of the quiver varieties is obtained 
\cite{KN, N2} essentially by 
replacing the bundles ${\underline V}, {\underline Q}\otimes {\underline V} 
\oplus {\underline W}$ 
appearing in (\ref{bundle_monad}) with bundles $(V \otimes {\cal R})^\Gamma$, 
$(Q\otimes V\otimes {\cal R})^\Gamma \oplus (W\otimes {\cal R})^\Gamma$, 
where ${\cal R}$ is a   
rank $|\Gamma|$ `tautological' bundle over $X_{\vec \xi}$, related to the  
description of $X_{\vec \xi}$ as a hyperkahler 
quotient. Then one lifts the operator $z\otimes$ 
to a section $\lambda$ of $Q\otimes End({\cal R})$. This lifts the 
monad (\ref{bundle_monad}) 
to the ALE space, allowing one to perform an ADHM 
construction over $X_{\vec \xi}$.

Our procedure above was entirely different in spirit, insamuch as we did 
not resolve the orbifold $Q/\Gamma$ `by hand'; rather, we deformed its algebra 
of smooth functions, which effectively resolved it via the uncertainty 
principle: since individual points cease to have a well-defined meaning, 
the orbifold singularity is lost in the noncommutative `fuzz'. 
The remarkable fact is that this apparently purely algebraic procedure gives 
a moduli space which has a classical geometric interpretation: for 
$V_0=0$, at least, instantons over the noncommutative {\em deformation} of 
the orbifold correspond to instantons over the {\em resolution} of the 
orbifold.

In \cite{NS} it was shown that Nakajima's resolution \cite{N} of the ideal 
instanton moduli space over $\R^4$ admits a noncommutative interpretation.  
The singularities of that space are due to instantons collapsing to zero 
size, and the heuristic reason for their smoothing out in the noncommutative 
set-up is that such a collapse is not possible once the fuzziness of the 
base space is taken into account. The moduli space of equivariant ideal 
instantons has a qualitatively new singularity 
due to instantons collapsing to zero size at the orbifold 
point. The noncommutative deformation of the orbifold  
effectively eliminates this point, thereby smoothing out this type of 
singularity as well. Therefore, the resolution mechanism is heuristically 
identical to that of \cite{NS}.

\section{Conclusions and further directions}

By studying the noncommutative version of the equivariant ADHM construction, 
we showed that the quiver varieties of \cite{N2} admit an interpretation
as moduli spaces of instantons over a noncommutative deformation of the 
Kleinian orbifold. In particular, the 
ALE spaces of \cite{K} can be described as a moduli space of 
$|\Gamma|-1$ $SU(2)$ 
noncommutative equivariant instantons.  These spaces also 
admit a description as moduli spaces of  equivariant torsion-free sheaves 
over $\C^2$, which lends further credence to  the conjectural correspondence 
\cite{NS} between noncommutative instantons and torsion-free sheaves.

There are two main lessons that we can learn from this, one of a physical  
and one of a mathematical nature. The physical lesson is that it is possible 
to describe moduli spaces of objects 
over the {\em resolution} of a singular space as moduli spaces of similar 
objects over a noncommutative {\em deformation} of that singular space. 
We believe that this  approach can play an important role in 
clarifying the nature of space-time resolution processes in matrix theory. 
The mathematical lesson is that rather 
nontrivial manifolds can be constructed as moduli spaces of noncommutative 
objects, and that classical geometric processes such as the minimal resolution 
of orbifold singularities admit a noncommutative geometric realization.

One point which we believe to be worth further study is the precise nature 
of the relation between torsion-free sheaves and noncommutative instantons. 
Both in \cite{NS} and in the present paper, such a 
correspondence was constructed at the level of moduli spaces only, 
but it is important to understand it at a more fundamental level.  
Namely, one would hope to have a relation between torsion-free sheaves 
and anti-self-dual noncommutative connections which is similar to the 
Hitchin-Kobayashi correspondence between stable holomorphic bundles and 
instantons. 
The best approach to this problem may be to view the noncommutative 
deformation of the base space as a certain type of {\em regularization}. 
Indeed, a generalized Hitchin-Kobayashi correspondence was constructed in 
\cite{BL}. This associates a singular  Einstein-Hermitian connection 
(of square-integrable curvature) to a polystable reflexive sheaf, and one 
may suppose that similar results could be derived for more general sheaves, 
if one replaces square-integrability with a weaker condition. We believe that 
the noncommutative deformation acts as a regulator of such connection 
singularities, which probably renders the curvature `square-integrable' in the 
noncommutative sense, similar to what happens in the examples considered in 
\cite{NS}. Such a `noncommutative' Hitchin-Kobayashi correspondence may 
shed considerable  light on the precise meaning of the relation between 
D-brane and instanton moduli spaces in string theory. 

Another interesting issue is the generalization of the previous results to the 
case of a compact base space, in particular to the case of orbifolds of 
noncommutative tori. This may help solve the problem of understanding the  
resolution of such singularities by D-branes. 
A direct approach to this problem leads to rather singular connections, 
which could be regularized by noncommutative-geometric methods. 

Finally, let us note that the procedure we used consists essentially of 
deforming the standard moment map to a noncommutative version, thereby 
absorbing nontrivial values of its level into the non-commutativity of the 
base space. Since symplectic quotients appear in the description of 
moduli spaces of various types of objects 
(as well as in the theory of Hamiltonian reduction of mechanical systems), it 
would very interesting to see whether a systematic 
deformation procedure exists in a 
more general set-up. Such a theory may contribute to a better understanding 
of the relation between classical and `quantum' geometry. 

\bigbreak\bigskip\bigskip\centerline{{\bf Aknowledgements}}\nobreak
\bigskip

The author wishes to thank N.~Nekrasov and 
A.~Schwarz for correspondence, I.~Kriechever for 
enlightening discussions and B.~R.~Greene for constant support and 
encouragement. 
This work was supported by a C.U. 
Pfister Fellowship and by the DOE grant DE-FG02-92ER40699B. 

\appendix

\section{Weyl quantization}

We start with the hermitian vector space $(Q,<,>)$ ( $<,>$ is 
anti-linear in the {\em first} variable). Let $(,):=Re<,>$ and 
$\omega(,):=Im<,>$ be the associated 
Euclidean scalar product and symplectic form on the underlying real vector 
space $Q_\R$. Fix $\eta \neq 0$. 
Quantization of the symplectic space $(Q_\R, \eta \omega)$ is 
achieved by considering a (strongly continuous) Weyl system 
$W_\eta:Q_\R\rightarrow {\cal U}({\cal H})$, with ${\cal H}$ a 
(separable) Hilbert space, satisfying the property:
\bdm
W_\eta(s)W_\eta(t)=W_\eta(s+t)e^{-\frac{i}{2}|\eta | \omega(s,t)} \nn
\edm
($s,t \in Q$). 
By von-Neumann's theorem, any such system is unitarily equivalent with 
the standard one, realized on the symmetric Fock space ${\cal H}$ over 
$(Q,<.>)$ by the operators(we use the formalism of 
\cite{Cook}, with the trivial deformation obtained by rescaling the 
creation/annihilation operators given there by $\sqrt{|\eta |}$):
\bdm
W_\eta(s):=e^{iQ_{|\eta |}(s)}~~\nn .
\edm
Here $Q_\eta(s):=\frac{1}{\sqrt{2}}[a^+_\eta(s)+a_\eta(s)]$ is the Segal 
operator with 
$a^+_\eta(s),a_\eta(s)$ the generation and annihilation operators of 
\cite{Cook}. 
The latter are $\C$-linear, respectively anti-linear in $s$ and satisfy:
\begin{eqnarray}
[a_\eta(s),a^+_\eta(t)]=|\eta |<s,t> ~,&
[a_\eta(s),a_\eta(t)]=0 ~,&
[a^+_\eta(s),a^+_\eta(t)]=0  \nn
\end{eqnarray}
for all $s,t \in Q$. The usual creation and 
annihilation operators $a^{\eta ~+}_i, a_i^\eta$ ($i=1,2$) are obtained by 
choosing an orthonormal basis ${\cal E}:=(e_i)_{i=1,2}$ of $(Q,<,>)$ and 
considering 
the associated real basis $(\epsilon_{i}:=e_i, \epsilon_{i+2}:=ie_i)_{i=1,2}$
of $Q_\R$. The latter is an orthonormal basis of $(Q_\R,(,))$ and a 
canonical (i.e. Liouville) basis of $(Q_\R,\omega)$. Then:
\begin{eqnarray}
a_i^\eta:=a_\eta(e_i) ~, & 
a^{\eta~+}_i:=a^+_\eta(e_i) ~~ \nn ,
\end{eqnarray}
while the usual momentum and position operators are :
\begin{eqnarray}
Q_i^\eta:=Q_\eta(\epsilon_i) ~, &~
P_i^\eta:=Q_\eta(\epsilon_{i+2})~~ \nn .
\end{eqnarray}
These have the nontrivial commutators:
\begin{eqnarray}
[a_i^\eta, a^{\eta~+}_j] =|\eta |\delta_{ij} ~, &
[Q_i^\eta, P_j^\eta]= i|\eta |\delta_{ij}~~ \nn .
\end{eqnarray}
The subspace ${\cal H}_0$ of states with a finite particle number is a dense 
subspace of the symmetric Fock space ${\cal H}$. 
In the position representation, we have 
${\cal H}\approx L^2(\R^2)$ and we define a dense 
subspace ${\cal H}_0$ of ${\cal H}$ by
${\cal H}_0\approx {\cal S}(\R^2)$. All 
operators we consider are densely defined on subspaces of ${\cal H}$ 
containing ${\cal H}_0$ and preserve ${\cal H}_0$. 

For any Schwarz function $f \in {\cal S}(Q_\R)$, define its Fourier transform 
by: 
\bdm
{\hat f}(s):=\frac{1}{(2\pi)^2}\int_{Q_\R}{dt~f(t)e^{i(s,t)}} \nn
\edm
where $dt$ is the natural Lesbegue measure on $(Q_\R,(,))$. 
Then Weyl quantization of the phase space distribution $f$ is achieved by the 
operator:
\be
\label{Weyl_transform}
W_f^\eta:=\frac{1}{(2\pi)^2}\int_{Q_\R}{dt~{\hat f_c}(-t)W_\eta(t)}~~.
\ee
where 
\bdm
f_c:=\left\{ \begin{array}{ll}
f, \mbox{if ~~} \eta > 0, \\
{\overline f}, \mbox{if ~~} \eta <0 
\end{array} \right .
\edm
This has the conjugation property:
\bdm
W_{f^*}^\eta=(W_f^\eta)^+~.
\edm
The Weyl correspondence (\ref{Weyl_transform}) can 
be extended to various spaces of functions and distributions on $Q_\R$ as 
explained in \cite{Folland, VB, Anderson}. 

If $f, g$ are such that $W^\eta_f, W^\eta_g$ are well-defined operators 
from ${\cal H}_0$ to ${\cal H}_0$, then the Weyl correspondence 
maps the operator product $W^\eta_fW^\eta_g$ into the Moyal product 
$f\times_\eta g$ via the relation:
\bdm
W_{f\times_\eta g}^\eta:=W_f^\eta W_g^\eta \nn
\edm
In particular, this product makes ${\cal S}(Q_\R)$ into an involutive algebra, 
which we denoted above by ${\cal A}_\eta$. 
As a set, ${\cal A}_\eta$ is independent of $\eta$. 
The $\eta$-dependence enters only through the product $\times_\eta$. 
For $f, g \in {\cal S}(Q_\R)$, the Moyal product is given by:
\bdm
f\times_\eta g= [e^{-\frac{i}{2}\eta\pi(d_u,d_v)}f(u)g(v)]_{u=v:=s} \nn
\edm
where $\pi$ is the symplectic form induced by $\omega$ on the dual space 
$(Q_\R)^*=Hom_\R(Q_\R, \R)$ and $d_u, d_v$ denote the total differentials 
$d_u(f):=d_uf \in Hom_\R(Q_\R,\R)$. This relation is valid for any sign of 
$\eta$. Note that taking $\eta$ into $-\eta$ amounts to replacing 
$f \times_\eta g$ by $f \times_{-\eta}g= g \times_{\eta} f$, i.e.  
`transposing' the algebra structure on ${\cal A}_\eta$. 

Choosing a basis ${\cal E}$ as before, one defines the real and complex 
coordinate functions: 
\begin{eqnarray}
x_i(s):=s^i ~, &~
x_{i+2}(s):=p_i(s):=s^{2+i} \nn \\
z_i:=x_i + ix_{i+2} ~, &~
{\overline z}_i:=x_i - ix_{i+2} \nn
\end{eqnarray}
where $s =\sum_{\alpha=1..4}{s^{\alpha}\epsilon_{\alpha}}=z^i(s)e_i 
\in Q$. 
Then
$\pi_{\alpha,\beta}:=\pi(\epsilon^*_\alpha,\epsilon^*_\beta)=-
\omega_{\alpha,\beta}:=-\omega(\epsilon_\alpha, \epsilon_\beta)$ and 
$df=\sum_{i=1,2}{(\frac{\partial{f}}{\partial{x_i}}+
\frac{\partial{f}}{\partial{p_i}})}$, where 
$(\epsilon^*_\alpha)_{\alpha =1..4}$ is the basis dual to 
$(\epsilon_\alpha)_{\alpha =1..4}$. 
This gives the standard form of the 
Moyal product:
\bdm
f\times_\eta g=fe^{\frac{i}{2}\eta \sum_{i=1,2}
{(\partial_{x_i}\partial_{p_i} -\partial_{p_i}\partial_{x_i})}}g~~ \nn .
\edm

The Weyl quantizations of $x_i,p_i$ are the operators 
$Q_i^\eta,P_i^\eta$, respectively. The Weyl quantizations of 
$(z_i, {\overline z}_i)$  are (in this order) 
$(\sqrt{2}a_i^\eta, \sqrt{2}a^{\eta~+}_i)$ for $\eta >0$ and  
$(\sqrt{2}a_i^{\eta~+}, \sqrt{2}a_i^\eta)$ for
$\eta<0$. These give the Moyal products (\ref{zs}) for all $\eta \neq 0$.

\

{\bf Action of phase-space translations and smooth vectors}

\

The Weyl group has an `adjoint' action by *-automorphisms of the  
$\C^*$-algebra  ${\cal B}({\cal H})$ of bounded operators in ${\cal H}$:
\bdm
B \rightarrow W(t)BW(t)^{-1}, ~\forall B \in {\cal B}({\cal H}) ~ \nn .
\edm
Let $B_c({\cal H})$, ${\cal B}^\infty({\cal H})$ be the subalgebras of 
continuous, respectively smooth vectors for this action. 
It was shown in \cite{Cordes, Payne} that $B_c({\cal H})$ is 
a $\C^*$- subalgebra of ${\cal B}({\cal H})$, ${\cal B}^\infty({\cal H})$ is a 
dense *-subalgebra of  $B_c({\cal H})$ and that the Weyl transform is a 
$*$-algebra isomorphism from ${\cal B}^\infty_\eta$ to 
${\cal B}^\infty ({\cal H})$. The adjoint action of the Weyl group is mapped 
into the action (\ref{trsl}) of $(Q_\R,+)$ on  
${\cal B}^\infty_\eta$:
\bdm
W_{\alpha_u(f)}^\eta=W_\eta(i|\eta |^{-1}u)W^\eta_fW_\eta(i|\eta |^{-1}u)^{-1} 
~~ \nn .
\edm

\

{\bf The action of the orbifold group}

\

For any $\gamma \in \Gamma$, let $U_\gamma$ be the `second-quantized' form 
of the unitary operator $\rho_Q(\gamma) \in U(Q,<,>)$ in our symmetric Fock 
space. $U_\gamma$ is independent of $\eta$ and satisfies:
\begin{eqnarray}
a_\eta(\rho_Q(\gamma)s)=U_\gamma a_\eta(s)U_\gamma^{-1} \nn \\
a^+_\eta(\rho_Q(\gamma)s)=U_\gamma a^+_\eta(s) U_\gamma^{-1}~~ \nn ,
\end{eqnarray}
which give:
\bdm
W_\eta(\rho_Q(\gamma)s)=U_\gamma W_\eta(s)U_\gamma^{-1}~~ \nn .
\edm
Then it is easy to see that:
\bdm
W^\eta_{f_\gamma}=U_\gamma W_f^\eta U_\gamma^{-1}~~ \nn ,
\edm
which implies that the transformations $\nu_\gamma:f\rightarrow f_\gamma$ 
are $*$-automorphisms of ${\cal B}^\infty_\eta$.

\subsection{The operator version of the ADHM construction} 

For what follows we take $\eta <0$. The Weyl transform 
maps the algebraic ADHM construction of section 2 into an operatorial 
construction, which we explain below. 

\

{\bf The operator $z$}

\

The Weyl quantization rule (\ref{Weyl_transform}) has a trivial generalization 
to vector-valued functions $s$ if we replace $f$ by $s$ in both sides. In this 
way, we can `quantize' any section of a trivial vector bundle 
${\underline S}$. In particular, we can define the Weyl transform 
${\hat z}_\eta:=W^\eta_z$ of the tautological section $z$ of ${\underline Q}$. 
In an orthonormal basis, this is given by:
\bdm
{\hat z}_\eta=\sqrt{2}a^{\eta~+}_i e_i={\hat z}^\eta_i e_i \nn
\edm
and is a (densely-defined) closed operator from ${\cal H}$ to 
$Q\otimes {\cal H}$. 

\

{\bf Operator form of ${\cal D}$, ${\cal D}^+$}

\

The Weyl quantizations ${\hat {\cal D}}_z, {\hat {\cal D}}_z^+$ of 
${\cal D}_z, {\cal D}_z^+$ (viewed as sections of the associated bundles 
of homomorphisms) are given by replacing $t$ with ${\hat z}_\eta$ in 
(\ref{D}).

If $S$ is a (finite dimensional) hermitian vector space, let ${\cal H}_S$ 
be the Hilbert space $S\otimes {\cal H}$ and ${\cal H}^0_S:=
S \otimes {\cal H}^0$ the associated dense subspace. Then 
${\hat {\cal D}}_z, {\hat {\cal D}}_z^+$ are densely defined operators 
between ${\cal H}_{V \oplus V }$ and ${\cal H}_{Q\otimes V \oplus W}$, whose 
domains contain ${\cal H}^0_{V \oplus V}$, respectively 
${\cal H}^0_{Q\otimes V \oplus W}$. They are closable on these subspaces, 
since ${\hat z}^\eta_i, {\hat z}^{\eta+}_i$ are closable. Denote their 
closures by the same letters. For any operator $A$, denote its domain 
by $d(A)$. 

\

{\bf The construction}

\

Consider the operator ${\hat {\cal D}_z}^+{\hat {\cal D}_z}$, defined 
on the set ${\cal M}=\{ x \in d({\hat {\cal D}}_z) | {\hat {\cal D}}_zx \in 
d({\hat {\cal D}}_z^+)\}$ (which includes ${\cal H}^0_{Q\otimes V \oplus W})$. 
It is well-known (see, for example, \cite{Simon_Reed}) that 
${\hat {\cal D}_z}^+{\hat {\cal D}_z}$ is self-adjoint on ${\cal M}$. 
By the ADHM equations, this operator has the form :
\bdm
{\hat {\cal D}_z}^+{\hat {\cal D}_z}=1_2 \otimes {\hat \Delta}= 
\left(\begin{array}{cc} 
{\hat \Delta} & 0 \\
0 & {\hat \Delta} \end{array}\right) \nn
\edm
with ${\hat \Delta}={\hat \sigma}^+_z{\hat \sigma}_z
={\hat \tau}_z {\hat \tau}^+_z$ a self-adjoint operator on 
$M =d({\hat \tau}_z {\hat \tau}^+_z)=d({\hat \sigma}^+_z{\hat \sigma}_z)
\subset {\cal H}_V$ 
(containing ${\cal H}^0_V$) with values in ${\cal H}_V$. The set 
${\cal M}$ equals $M \oplus M$. Assuming as in \cite{NS} 
that $\Delta$ is (strictly) 
positive, its inverse $\Delta^{-1}$ is a bijective bounded operator from 
${\cal H}_V$ to M. We further assume that 
$\Delta^{-1}({\cal H}^0_V)\subset {\cal H}^0_V$.

\

{\em The projector ${\cal P}$} 

\

Let $K:=ker {\hat {\cal D}}_z^+$ 
and $P$ be the orthogonal projector on $K$ in 
${\cal H}_{Q\otimes V \oplus W}$. 
Then it is not hard to see that 
(with the assumptions above):
\bdm
P|_{d({\cal D}^+_z)}=1 - {\cal D}_z(1_2 \otimes \Delta^{-1}) {\cal D}^+_z ~ 
\nn .
\edm

For any (finite -dimensional) hermitian vector space $S$, consider the 
space ${\cal B}_S:= {\cal B}({\cal H} , {\cal H}_S) =
S \otimes {\cal B}({\cal H})$ of bounded operators from ${\cal H}$ to 
${\cal H}_S$. 
This has a natural hermitian ${\cal B}({\cal H})$- module structure given 
by $<A,B>:=A^+B$. 
We also let ${\cal A}_S, {\cal B}^\infty_S$  be the subsets of operators in 
${\cal B}_S$ whose 
Weyl symbols belong to $S\otimes{\cal S}(Q_\R)$, $S\otimes{\cal B}^\infty_0$, 
both viewed as ${\cal B}^\infty({\cal H})$-modules. ${\cal A}_S$ is a 
${\cal B}({\cal H})$-submodule of ${\cal B}_S$. The Weyl transform 
(acting on both the module and the base algebra) gives a
unitary isomorphism between the modules  
$S\otimes {\cal B}_\eta^\infty$, $S\otimes {\cal A}_\eta$ and
${\cal B}_S^\infty$, ${\cal A}_S$ respectively.

$P$ induces an orthogonal projector ${\cal P}$ of the 
hermitian module ${\cal B}_{Q\otimes V \oplus W}$ via:
\bdm
{\cal P}(A):=PA , ~\forall A \in  {\cal B}_{Q\otimes V \oplus W}~~ \nn .
\edm
${\cal P}$ projects onto the subspace ${\cal E}_b$ of all 
$A \in {\cal B}_{Q\otimes V \oplus W}$ whose range is a subset of $K$. 
If $A \in  {\cal B}_{Q\otimes V \oplus W}$ preserves ${\cal H}_0$, then 
clearly $A \in {\cal E}_b$ iff ${\hat {\cal D}}^+_zA|_{{\cal H}_0}=0$, 
which is equivalent to the fact that ${\cal D}^+$ annihilates 
the Weyl symbol of $A$. 

Now further assume that ${\cal P}$ preserves  
${\cal A}_{Q\otimes V \oplus W}$. In this case, let 
${\cal E}:={\cal P}({\cal A}_{Q\otimes V \oplus W})$. Then the Weyl transform 
maps ${\cal E}$ to the ${\cal B}_\eta^\infty$-module ${\cal E}_\eta$
and ${\cal P}$ to the projector ${\cal P}_\eta$ used in section 2. 

\

{\em The map $\psi$}

\

Choosing an orthonormal 
basis of $Q$ for simplicity, we can write:
\bdm
{\hat {\cal D}}^+_z=({\hat D}_z^+, V) \nn ~~, 
\edm
where ${\hat D}_z^+$ is the  closed densely-defined operator from 
${\cal H}_{Q\otimes V}$ to ${\cal H}_{V \oplus V}$ given by:
\bdm
{\hat D}_z^+:=\left(\begin{array}{cc}
-B_2+z_\eta^2 & B_1 -z_\eta^1 \\
B_1^+-z_\eta^{1 ~+} & B_2^+-z_\eta^{2~+} 
\end{array}\right)~~ \nn ,
\edm
while $V$ is the bounded operator from 
${\cal H}_W$ to ${\cal H}_{V \oplus V}$ given by:
\bdm
V:=\left(\begin{array}{c}
i\\
j^+ 
\end{array}\right)~~ \nn .
\edm
The domain of ${\hat {\cal D}}_z^+$ is the direct sum of the  
domain of  ${\hat D}_z^+$ with the entire space $W \otimes {\cal H}$. 
Assuming as in \cite{NS} that $0$ does not belong to the spectrum of 
${\hat D}_z$
we have a bounded inverse 
$R_0:= ({\hat D}_z^+)^{-1}$ defined on ${\cal H}_{V \oplus V}$. 
Defining a bounded operator 
$S:=R_0 V:{\cal H}_W\rightarrow {\cal H}_{Q\otimes V}$, 
the kernel $K$ of ${\hat {\cal D}}_z^+$ coincides with the graph of 
$-S$. Define $\Psi_0:=(-S) \oplus id_{{\cal H}_W}:{\cal H}_W \rightarrow 
{\cal H}_{Q\otimes V \oplus W}$ and let 
$\Psi:=(\Psi_0^+ \Psi_0)^{-1/2}\Psi_0$ be its unitary part. 
Then $\Psi$ induces a map of hermitian modules 
$\psi:{\cal B}_W \rightarrow {\cal B}_{Q\otimes V \oplus W}$ by 
\bdm
\psi(A):=\Psi A,~\forall A \in {\cal B}_W~~ \nn .
\edm
Since $im \Psi =K$, this gives a unitary isomorphism between the hermitian 
modules ${\cal B}_W$ and ${\cal E}_b$. Assuming that 
$\psi({\cal A}_W)={\cal E}$, the inverse Weyl transform 
gives the map used in section 2.

\

{\bf Extension to ${\cal E}^\infty$ }

\

It is clear from the above that we can extend the whole discussion of the 
ADHM construction to the modules $V \otimes {\cal B}^\infty$, 
$(Q\otimes V\otimes {\cal B}^\infty)$ etc. provided 
that $P$ satisfies a slightly different condition, namely 
that left multiplication by $P$ maps 
${\cal B}^\infty_{Q\otimes V \oplus W}$ into itself. 
In this case, we can define ${\cal E}^\infty={\cal P}({\cal E}_b)=
\{A \in {\cal B}^\infty_{Q\otimes V \oplus W}|im A \in K\}$.
Then the inverse Weyl transform of $P$ gives the projector 
${\cal P}_\eta^\infty$ of section 3, while the inverse Weyl transform 
of ${\cal E}^\infty$ gives the module ${\cal E}_\eta^\infty$. 
Despite its apparent 
simplicity, this `regularity' condition on $P$ is nontrivial. 
It could be tested, in principle, by 
computing the anti-Wick symbol of $P$ and using the results of \cite{Berezin}.

\section{Quiver varieties, instantons over ALE spaces and the  resolution of 
the moduli space of orbifold instantons}

\

{\bf Definitions and basic results}

\

One starts with equivariant ADHM data as in section 2. 
Let ${\bf g}_V$ be the Lie algebra of $G_V$ 
and ${\cal Z}_V$ be its center. 
By the decomposition $V=\oplus_{i=0..r}{V_i \otimes R_i}$, we have 
$G_V=\Pi_{i=0..r}{U(V_i)}$, ${\bf g}_V=\Pi_{i=0..r}{u(V_i)}$ and 
${\cal Z}_V$ is the subset of skew-hermitian operators which are diagonal on 
each $V_i$. We identify ${\cal Z}_V$ with the set:
\bdm
Z_V:=\{\zeta=(\zeta_0...\zeta_r)\in \R^{r+1}|\zeta_i=0\  \rm{if}\  V_i=0\} \subset 
\R^{r+1} \nn 
\edm 
via the map $\zeta \in Z_V\rightarrow {\hat \zeta}:=\frac{i}{2}
\oplus_{k=0..r}{\zeta_k 1_{V_k}}$. 
As before, we have a hyperkahler moment map 
${\vec \mu} :P_\Gamma \rightarrow \R^3 \otimes {\bf g}_V$ or, equivalently, a 
real and a complex moment map $\mu_r=\mu_1$, $\mu_c=\mu_2+i\mu_3$. 

The resolution of the equivariant instanton moduli space can be achieved 
by replacing the usual ADHM equations with their inhomogeneous version. 
For this, let ${\vec \zeta} \in \R^3 \otimes Z_V$. Define the following 
{\em quiver varieties} \cite{N2}: 
\bdm
{\cal M}_{\vec \zeta} (V,W):=\{ ({\cal B},i,j) \in P_\Gamma(V,W) 
| {\vec \mu}({\cal B},i,j)=-{\hat {\vec \zeta}} \} /G_V \nn
\edm

\bdm
{\cal M}^{reg}_{\vec \zeta} (V,W):=\{({\cal B},i,j) \in P^{reg}_\Gamma(V,W) |  
{\vec \mu}
({\cal B},i,j)=-{\hat {\vec \zeta}} \}/G_V \subset {\cal M}_{\vec \zeta}
(V,W)~~ \nn .
\edm
In general, ${\cal M}_{\vec \zeta}(V,W)$ has complicated singularities, but 
${\cal M}^{reg}_{\vec \zeta}(V,W)$ is a smooth hyperkahler manifold, of dimension 
$d(v,w)=4{\vec v}{\vec w} -2 {\vec v}C{\vec v}$.

To give a sufficient criterion of smoothness of ${\cal M}_{\vec \zeta}(V,W)$, 
one defines: 

$R_+:=\{\theta=(\theta_0..\theta_r) \in (\Z_{\geq 0})^{r+1}|
\theta^t C \theta \leq 2\}-\{0\}$

$R_+({\vec v}):=
\{\theta \in R_+| \theta_k \leq v_k, \forall k=0..r\}\subset R_+$

\noindent and, for all $\theta \in R_+$:

$D_\theta:=\{x=(x_0..x_r) \in \R^{r+1} | \sum_{k=0..r}{x_k \theta_k} =0\}$.

\noindent We say that ${\vec \zeta} \in \R^3\otimes Z_V$ is {\em generic} if 
${\vec \zeta}$ does not belong to $\cup_{\theta \in R_+({\vec v})}{\R^3 \otimes 
D_\theta}$. 
If ${\vec \zeta}$ is generic, then it is shown in \cite{N2} that:

(1)${\cal M}_{\vec \zeta}(V,W)={\cal M}^{reg}_{\vec \zeta}(V,W)$. Thus ${\cal M}_{\vec \zeta}(V,W)$ 
is smooth; moreover, its hyperkahler metric is complete. 

(2)Assuming further that ${\cal M}^{reg}_{(0,\zeta_c)}(V,W)$ is nonempty, 
then there exists a map 
${\cal M}_{\vec \zeta}(V,W) \rightarrow {\cal M}_{(0,\zeta_c)}(V,W)$ which 
is a resolution of singularities. 
In particular, given $\zeta_r\in Z_V$ such that  $(\zeta_r,0)$ is generic,  
we have a resolution ${\cal M}_{(\zeta_r,0)}(V,W) \rightarrow 
{\cal M}_0(V,W)$.

For any  ${\vec \zeta}$, 
$\vec{\zeta}'$ which 
are generic, the manifolds ${\cal M}_{\vec \zeta}(V,W)$ and 
${\cal M}_{\vec \zeta}'(V,W)$  
are diffeomorphic. Hence given a generic $(\zeta_r,0)$, the manifold 
${\cal M}_{(\zeta_r,0)}$ provides a differential model for all 
${\cal M}_{\vec \zeta}(V,W)$ with a generic ${\vec \zeta}$.

\

{\bf Instanton interpretation}

\

If the parameters ${\vec \zeta} \in \R^3 \otimes Z_V$ have a particular form, 
then ${\cal M}^{reg}_{\vec \zeta}(V,W)$ admits an interpretation as 
a moduli space of instantons over an ALE space, while  
${\cal M}_{\vec \zeta}(V,W)$ coincides with the associated moduli space of 
{\em ideal} instantons. To explain this, we first describe Kronheimer's 
construction of ALE spaces as hyperkahler quotients.

\

{\em Kronheimer's construction of ALE spaces}

\

According to \cite{K}, all 4-dimensional ALE spaces can be obtained as 
hyperkahler quotients. For this, consider $V=R$ (the regular 
representation of $\Gamma$) and $W=0$. In this rather degenerate case, 
the central $U(1)$ subgroup of $G_R=U_\Gamma(R)$ acts trivially on 
$P^{reg}_\Gamma(V,W)$. To overcome this pathology, one replaces $G_R$ by the 
quotient group $G'_R:=G_R/U(1)$, which acts freely.  
The Lie algebra ${\bf g}'_R$ of $G'_R$ is the traceless part of the 
Lie algebra of $G_R$. Its center ${\cal Z}'_R$ is formed of $r \times r$ 
{\em traceless} diagonal matrices, with eigenvalues occurring in blocks 
of dimensions $n_j$ ($j=0..r$). This can be identified with:
\bdm
Z'_R:=\{(\xi_0..\xi_r)| \sum_{i=0..r}{n_i\xi_i}=0\}~\subset \R^{r+1} \nn
\edm
via the map $\xi=(\xi_0..\xi_r)\rightarrow {\hat \xi}:=\frac{i}{2}
\oplus_{k=0..r}{\xi_k 1_{V_k}}$. 
One chooses ${\vec \xi} \in \R^3 \otimes Z'_R$ and defines: 
\bdm
X_{\vec \xi}:=\{({\cal B},i,j) \in P_\Gamma(R,0)|{\vec \mu}({\cal B},i,j)=
+{\hat {\vec \xi}}\}/G'_R \nn
\edm
\bdm
X^{reg}_{\vec \xi}:=\{({\cal B},i,j) \in P^{reg}_\Gamma(R,0)|
{\vec \mu}({\cal B},i,j)=+{\hat {\vec \xi}}\}/G'_R~~ \nn.
\edm
The parameter ${\vec \xi}$ is called {\em non-degenerate} if 
${\vec \xi} \in \R^3 \otimes Z'_R - \cup_{\theta \in R_+ 
-\{q{\vec n}|q \in \Z\}}\R^3\otimes {D_\theta}$.

If ${\vec \xi}$ is non-degenerate, then one has \cite{KN} 
$X_{\vec \xi}=X^{reg}_{\vec \xi}$ and $X_{\vec \xi}$ is a smooth ALE 
hyperkahler manifold of real 
dimension 4, which is diffeomorphic with the minimal resolution of the 
Kleinian singularity $Q/\Gamma$. 
For ${\vec \xi}=0$, one has $X_0=Q/\Gamma$.  

A more useful form of the non-degeneracy condition can be obtained as follows. 
Consider the set $\Phi:=\{{\tilde \theta}\in \Z^r|\sum_{i,j=1..r}
{C_{ij}{\tilde \theta}_i{\tilde \theta}_j}= 2\}$, which can be identified with 
the set of simple roots of ${\bf g}_\Gamma$. Then  
$(\theta_0..\theta_r)\rightarrow 
(\theta_1-n_1\theta_0,..,\theta_r-n_r\theta_0)$ maps the set $R_+^0:=R_+ 
-\{q{\vec n}|q \in \Z\}$ {\em onto} $\Phi$, while the map 
$(\xi_1..\xi_r)\rightarrow (-\sum_{k=1..r}{n_k\xi_k}, \xi_1..\xi_r)$ 
is a bijection from $\R^r$ to $Z'_R$. Using this, is is easy to see that 
${\vec \xi} \in \R^3 \otimes Z'_R$ is non-degenerate iff 
$\sum_{i=1..r}{{\tilde \theta}_i{\vec \xi_i}}\neq 0, 
~\forall {\tilde \theta} \in \Phi$. 

$X_{\vec \xi}$ can be presented as a quiver variety as follows. Choose 
$V=R\ominus R_0$ and $W=Q$. Then $R_+(R\ominus R_0)\subset R_+^0$ can be 
identified with $\Phi$ in the obvious way and it is clear that 
${\vec \xi} \in Z'_R$ is non-degenerate iff $\phi({\vec \xi})=
({\vec \xi}_1..{\vec \xi}_r)$ is generic. 
In this case, it is easy to see that \cite{N1, KN}:
\bdm
X_{\vec \xi}={\cal M}_{\phi({\vec \xi})}(R\ominus R_0, Q)~~ \nn .
\edm
Moreover, by the results  we discuss below, 
${\cal M}_{\phi({\vec \xi})}(R\ominus R_0, Q)$
coincides with the moduli space of instantons over $X_{\vec \xi}$
(framed at infinity) of topological invariants 
$c_1=0, c_2[X_{\vec \xi}]=\frac{dim V}{|\Gamma|}$ 
and isotropy representation at infinity given by $Q$.

\

{\em Instanton description}

\

Define a map $\phi:Z'_R \rightarrow Z_V$ by :

\bdm
\phi(\xi_0...\xi_r):=(s_0\xi_0....s_r\xi_r) \nn
\edm
where $s_i:=0$ or $1$ according to whether $\dim_\C V_i=0$ or $\neq 0$, and  
let $Z_V^\phi$ be the image of this map in $Z_V$. In general, $Z_V^\phi$ is 
strictly smaller than $Z_V$. The desired interpretation is possible 
only if ${\vec \zeta} \in \R^3 \otimes Z^\phi_V$. More precisely, let 
${\vec \xi} \in \R^3 \otimes Z'_R$ be {\em non-degenerate} and let 
${\vec \zeta}:=\phi({\vec \xi})$. 
Then we have the following statements \cite{KN,N2}:

(a)${\cal M}^{reg}_{\vec \zeta}(V,W)$ coincides with the moduli space 
of instantons (framed at infinity) over $X_{\vec \xi}$ 
(of Chern classes and isotropy representation 
at infinity determined by $V,W$ as explained in \cite{KN});

(b)${\cal M}_{\vec \zeta}(V,W)$ coincides with the associated moduli space 
of {\em ideal} instantons. 

\noindent Note that :

($\alpha$) If $V_i \neq 0$ for at least one $i\in \{0..r\}$, then 
$Z^\phi_V=Z_V$, i.e. $\phi$ is surjective; in this case, $Z^\phi_V=Z_V$ 
and we have an instanton 
interpretation for all ${\vec \zeta}$ associated via $\phi$ with a 
{\em non-degenerate } $\xi \in Z'_R$.  

($\beta$)If $V_i \neq 0$ for all $i=0..r$, then 
$Z^\phi_V=\{\zeta=(\zeta_1..\zeta_r) 
\in \R^{r+1}|\sum_{i=0..r}{n_i\zeta_i}=0\}$, 
while $Z_V=\R^{r+1}$. In particular, $\phi$ is not surjective and $Z_V^\phi$ 
is strictly smaller than $Z_V$.

\

{\bf Diagonal levels of the real moment map}

\

Given a real parameter $\zeta$, consider the moment map level 
$\mu_r=-\frac{i}{2}\zeta 1_V, \mu_c=0$, which is represented by the 
parameters $\zeta_r=(s_0...s_r)\zeta, \zeta_c=0$. It is immediate that 
$(\zeta_r,0)$ is always generic, so that ${\cal M}_{(\zeta,0)}(V,W):=
{\cal M}_{(\zeta_r,0)}(V,W)$ is always smooth. Now consider the ALE 
instanton interpretation for this space:

(A)If $V_i\neq 0, \forall i=0..r$, then 
all $s_i$ are strictly positive and $\zeta_r$ never belongs to 
$Z_V^\phi$. Therefore, an ALE instanton interpretation is not possible.

(B)On the other hand, if $V_i=0$ for at least one $i$
(in which case $Z^\phi_V=Z_V$), and 
if $(\zeta_r,0) =\phi({\vec \xi}_r)$, with a {\em non-degenerate} 
${\vec \xi_r}$, then the {\em hyperkahler} manifold 
${\cal M}_{(\zeta_r,0)}(V,W)$ coincides with 
the moduli space of instantons over the smooth ALE space $X_{{\vec \xi}_r}$. 

To see whether $(\zeta_r,0)$ can be represented by a non-degenerate 
${\vec \xi}_r$,
consider the most restrictive case, namely when $V_m=0$ for some $m \in 
\{0..r\}$ and $V_k\neq 0, ~\forall k \neq m$. In this case, the equation 
$\phi({\vec \xi})=(\zeta_r,0)$ has exactly one solution, 
namely $\xi^c=0$ and 
$\xi^r_m=-\frac{N-n_m}{n_m}\zeta$, $\xi^r_i=\zeta, ~\forall i\neq m$, where 
$N:=\sum_{k=0..r}{n_k}$. It is easy to see that such a $(\xi_r,0)$ is in 
general non-degenerate, 
but we do not have a uniform proof that this 
holds for any group $\Gamma$. The case $V_0=0$, $V_i\neq 0, ~\forall i=1..r$ 
is particularly simple, since in this situation we have 
$\sum_{i=1..r}{\xi^r_i{\tilde \theta}_i}=
\zeta\sum_{i=1..r}{{\tilde \theta}_i}$, 
which is nonzero for any ${\tilde \theta} \in \Phi$. 

Thus $(\xi_r,0)$ is non-degenerate for any $\Gamma$ and any $V$ such 
that $V_0=0$, and in this case, at least, one can 
represent the hyperkahler manifold 
${\cal M}_{(\zeta_r,0)}(V,W)$ as a moduli space of instantons over 
the smooth ALE space $X_{(\xi_r,0)}$.

\

{\bf Sheaf-theoretic interpretation of ${\cal M}_{(\zeta,0)}(V,W)$ }

\

The sheaf -theoretic interpretation of ${\cal M}_{(\zeta,0)}(V,W)$ can be 
obtained essentially as follows. Consider the (`commutative') 
sequence (\ref{bundle_monad}) for data satisfying the homogeneous 
complex ADHM equation.
By virtue of this equation, (\ref{bundle_monad})  
is still a complex: $\tau\sigma=0$. 
Replacing the bundles by their sheaves of holomorphic 
sections, we can pull this back to a sheaf complex over $\C\P^2$ via the 
usual projection $p:\C\P^2\rightarrow {\overline Q}$. $p$ takes the line 
$l_\infty$ at infinity in $\C\P^2$ into the point $\infty \in {\overline Q}$, 
so the resulting sheaves over $\C\P^2$ are trivial over $l_\infty$. Moreover, 
$p^*$ takes any framing over $\infty$ into a framing over $l_\infty$. 
An equivariant version of the argument in \cite{N_lectures} 
shows that the first map 
of the resulting complex is always injective (as a {\em sheaf} map), while 
the second map is surjective iff condition (b) is satisfied, which is 
in turn equivalent (using the 
`transpose' form of Proposition 3.5. of \cite{N2}) with the requirement that 
the $G_V^{\C}$-orbit of $({\cal B},i,j)$ intersects the level 
$\mu_r=-\frac{i}{2}\zeta Id_V$ (the fact that $\zeta <0$ is crucial at this 
point). 
Thus, we have a 
sheaf {\em monad} over $\C\P^2$ iff there exists a $g \in G_V^{\C}$ such that
\bdm
\mu_r(g{\cal B}_1g^{-1}, g{\cal B}_2g^{-1}, gi, jg^{-1})=
-\frac{i}{2}\zeta Id_V~~ \nn . 
\edm
Moreover, such a $g$ is determined up to 
multiplication on the left by an element of $G_V$. 
The cohomology of the $\C\P^2$ monad gives a sheaf, which 
carries an induced action of $\Gamma$.  Repeating the arguments of 
\cite{OS, N_lectures} in this context, the 
quotient:
 
$\{({\cal B}, i,j)|\mu_c({\cal B},i,j)=0 $ and  
$G_V^{\C}({\cal B}, i,j) \cap \mu_r^{-1}(-\frac{i}{2}\zeta Id_V)\neq \Phi \}/
(G_V)^{\C}$ 

\noindent gives a moduli space of $\Gamma$-equivariant torsion free sheaves 
(framed over $l_\infty$) over $\C\P^2$, 
which can be identified with the {\em complex} manifold 
${\cal M}_{(\zeta,0)}(V,W)$.

\end{document}